\documentclass[12pt]{article}
\usepackage{latexsym, epsfig, amssymb, amsmath}
\usepackage[dvips]{color}
\usepackage{graphicx}
\usepackage{anysize}

\makeatletter

\renewcommand{\baselinestretch}{1.6}
\marginsize{1.0 in}{1.0 in}{1.25 in}{1.25 in}

\renewcommand \theequation{\arabic{section}.\arabic{equation}}

\def\singlespace{\def\baselinestretch{1}\@normalsize}

\newtheorem{theorem}{Theorem}

\newtheorem{definition}[theorem]{Definition}

\newtheorem{lemma}{Lemma}

\def\1{{\mathbf 1}}        % indicator
\def\var{{\mathop{\mathbf Var}}}    % variance
\def\T{{\mathbf T}}
\def\D{{\mathbf D}}
\def\H{\mathbf{H}}
\def\y{\mathbf{ y}}
\def\0{\mathbf{ 0}}
\def\C{\mathbf{C}}
\def\W{\mathbf{W}}
\def\w{\mathbf{w}}
\def\s{\mathbf{s}}
\def\G{\mathbf{G}}
\def\z{\mathbf{z}}

\def\b{\mathbf{b}}
\def\e{\mathbf{e}}

\def\v{{\mathbf{v}}}
\def\X{{\mathbf{X}}}

\def\balpha{\boldsymbol{\alpha}}

\def \S{\mathbf{S}}

\def \etal{{\em et al. }}
\def\bSigma{\boldsymbol{\Sigma}}
\def\bOmega{\boldsymbol{\Omega}}
\def\bphi{\boldsymbol{\phi}}
\def\boldeta{\boldsymbol{\eta}}
\def\bphinhat{\boldsymbol{\hat{\phi}}_n}

\def\bphin{\boldsymbol{\phi}_{n}}
\def\phihat{\hat{\phi}}
\def\bl{\boldsymbol{\ell}}

\def\bmu{\boldsymbol{\mu}}
\def\1{\mathbf{1}}
\def\0{\mathbf{0}}

\def\bzeta{\boldsymbol{\zeta}}

\def\u{\mathbf{u}}

\def\tr{\text{tr}}

\def\blj0{\bl_j^{(0)}}

\def\diag{\text{diag}}
\def\var{\text{var}}
\def\bepsilon{\boldsymbol{\epsilon}}

\def\dcv{\stackrel{\scriptscriptstyle \mathcal{D}}{\longrightarrow}}        % convergnece in distribution
\makeatletter
\def\table{\@ifnextchar[{\table@i}{\table@i[\fps@table]}}
\def\table@i[#1]{\@float{table}[#1]\footnotesize}
\makeatother

\begin{document}

\title{\vspace{-0.9 in} \bf Estimation of Large Precision
Matrices Through Block Penalization
\footnote{
Clifford Lam, PhD student
(Email: wlam@princeton.edu. Phone: (609) 240-6928).
Financial support from the NSF grant DMS-0704337 and NIH grant
R01-GM072611 is gratefully acknowledged.
}
\date{}
\author{By Clifford Lam \\
Department of Operations Research and Financial Engineering \\
Princeton University, Princeton, NJ, 08544}} \maketitle

\vspace{-0.25 in}
\begin{singlespace}
\begin{quotation}
\indent

This paper focuses on exploring the sparsity of the inverse
covariance matrix $\bSigma^{-1}$, or the precision matrix.
We form blocks of parameters based on each
off-diagonal band of the Cholesky factor from its modified Cholesky
decomposition, and penalize each block of parameters using the
$L_2$-norm instead of individual elements. We
develop a one-step estimator, and prove an oracle property which consists of a notion of
block sign-consistency and asymptotic normality. In particular, provided the initial
estimator of the Cholesky factor is good enough and the true Cholesky has finite number of non-zero off-diagonal bands, oracle property holds
for the one-step estimator even if $p_n \gg n$, and can even be as large as $\log p_n = o(n)$,
where the data $\y$ has mean zero and tail
probability $P(|y_j| > x) \leq K\exp(-Cx^d)$, $d > 0$, and $p_n$ is the number of variables. We also prove an
operator norm convergence result,
showing the cost of dimensionality is just $\log p_n$. The
advantage of this method over banding by Bickel and Levina (2008)
or nested LASSO by Levina \emph{et al.} (2007) is that it allows
for elimination of weaker signals that precede stronger ones in the
Cholesky factor. A method for obtaining an initial estimator for
the Cholesky factor is discussed, and a gradient projection
algorithm is developed for calculating the one-step estimate.
Simulation results are in favor of the newly proposed method and a set of real data is analyzed
using the new procedure and the banding method.

\end{quotation}
\end{singlespace}

{\em Short Title}: Block-penalized Precision Matrix Estimation.



\section{Introduction}\label{sect:introduction}

The need for estimating large covariance matrices arises naturally
in many scientific applications. For example in bioinformatics,
clustering of genes using genes expression data in a microarray
experiment; or in finance, when seeking a mean-variance efficient
portfolio from a universe of stocks. One common feature is that
the dimension of the data $p_n$ is usually large compare with the
sample size $n$, or even $p_n \gg n$ (genes expression data, fMRI
data, financial data, among many others). The sample covariance matrix $\S$ is
well-known to be ill-conditioned in such cases. Even for $\bSigma
= I$ the identity matrix, the eigenvalues of $\S$ are more spread
out around 1 asymptotically as $p_n/n$ gets larger (the
Mar\^{c}enko-Pastur law, Mar\^{c}enko and Pastur, 1967). It is singular when $p_n >
n$, thus not allowing an estimate of the inverse of the covariance
matrix, which is needed in many multivariate statistical
procedures like the linear discriminant analysis (LDA), regression
for multivariate normal data, Gaussian graphical models or portfolio allocations.  Hence
alternatives are needed for more accurate and useful estimation of covariance
matrix.

One regularization approach is penalization, which is the main
focus of this paper. Sparse estimation of the precision matrix
$\bOmega = \bSigma^{-1}$ has been investigated by many
researchers, which is very useful in Gaussian graphical models or
covariance selection for naturally ordered data (e.g. longitudinal
data, see Diggle and Verbyla (1998)). Meinshausen and B\"{u}hlmann
(2006) used the $L_1$-penalized likelihood to choose suitable
neighborhood for a Gaussian graph and showed that $p_n$ can grow
arbitrarily fast with $n$ for consistent estimation, while Li and
Gui (2006) considered updating the off-diagonal elements of
$\bOmega$ by penalizing on the negative gradient of the
log-likelihood with respect to these elements. Banerjee, d'Aspremont
and El Ghaoui (2006) and Yuan and Lin (2007) used $L_1$-penalty to
directly penalize on the elements of $\bOmega$, and develop
different semi-definite programming algorithms to achieve sparsity
of the inverse. Friedman, Hastie and Tibshirani (2007) and
Rothman \etal (2007) considered maximizing the $L_1$-penalized Gaussian
log-likelihood on the off-diagonal elements of the precision
matrix $\bOmega$, where the Graphical LASSO and the SPICE
algorithms are proposed respectively in their papers for finding a
solution, and the latter proved Frobenius and operator norms
convergence results for the final estimators.

Pourahmadi (1999) proposed the modified Cholesky decomposition
(MCD) which facilitates greatly the sparse estimation of $\bOmega$
through penalization. The idea is to decompose $\bSigma$ such that
for zero-mean data $\y = (y_1,\cdots,y_{p_n})^T$, we have for
$i=2,\cdots,p_n$,
\begin{equation}\label{eqn:MCD}
y_i = \sum_{j=1}^{i-1} \phi_{i,j}y_j + \epsilon_i, \text{ and }
\T\bSigma\T^T = \D,
\end{equation}
where $\T$ is the unique unit lower triangular matrix with ones on
its diagonal and $(i,j)^\text{th}$ element $-\phi_{i,j}$ for $j <
i$, and $\D$ is diagonal with $i^\text{th}$ element $\sigma_i^2 =
\text{var}(\epsilon_i)$. The optimization problem is unconstrained
(since the $\phi_{ij}$'s are free variables), and the estimate for
$\bOmega$ is always positive-definite. With MCD in
(\ref{eqn:MCD}), Huang \emph{et al.} (2006) used the $L_1$-penalty
on the $\phi_{i,j}$'s and optimized a penalized Gaussian
log-likelihood through a proposed iterative scheme, with the case
$p_n < n$ considered. Levina, Rothman and Zhu (2007) proposed a novel
penalty called the nested LASSO to achieve a flexible banded
structure of $\T$, and demonstrated by simulations that normality
of data is not necessary, with $p_n > n$ considered.

%\subsection{An example: the need for alternative methodology}\label{subsect:BPexample}

For estimating the precision matrix $\bOmega$ for naturally
ordered data, apart from the nested LASSO, Bickel and Levina
(2008) proposed banding the Cholesky factor $\T$ in
(\ref{eqn:MCD}), with the banding order $k$ chosen by minimizing a
resampling-based estimation of a suitable risk measure. The method
works on estimating a covariance matrix as well. While these two
methods are simple to use, they cannot eliminate blocks of weak
signals in between stronger signals. For instance, consider a time
series model
\begin{equation*}
y_i = 0.7y_{i-1} + 0.3y_{i-3} + \epsilon_i,
\end{equation*}
which corresponds to (\ref{eqn:MCD}) with $\phi_{i,2} = 0$,
$\phi_{i,j} = 0$ for $j \geq 4$. For example, this kind of model
can arise in clinical trials data, where response on a drug for
patients follows a certain kind of autoregressive
process with weak signals preceding stronger ones. This implies a
banded Cholesky factor $\T$, with the first and third
off-diagonal bands being non-zero and zero otherwise. Banding and
nested LASSO can band the Cholesky factor $\T$ starting from the
fourth off-diagonal band, but cannot set the second off-diagonal band to
zero. And if these methods choose to set the second off-diagonal band
to zero, then the third non-zero off-diagonal band will be wrongly set
to zero. Both failures can lead to
inaccurate analysis or prediction, in particular the maximum
eigenvalue of a precision matrix can then be estimated very wrongly. Clearly, an alternative method is required in
this situation. We present the block penalization framework in the
next section and more motivations and details of the methodology.

For more references, Smith and Kohn (2002) used a hierarchical
Bayesian model to identify the zeros in the Cholesky factor $\T$
of the MCD. Fan, Fan and Lv (2007), using factor analysis,
developed high-dimensional estimators for both $\bSigma$ and
$\bSigma^{-1}$. Wu and Pourahmadi (2003) proposed a banded
estimator through smoothing of the lower off-diagonal bands of
$\hat{\T}$ obtained from the sample covariance matrix (implicitly,
$p_n < n$). Then an order for
banding of $\hat{\T}$ is chosen by using AIC penalty of normal
likelihood of data. Furrer and Bengtsson (2007) considered
gradually shrinking the off-diagonal bands' elements of the sample
covariance matrix towards zero. Bickel and Levina (2007) and El
Karoui (2007) proposed the use of entry-wise thresholding to
achieve sparsity in covariance matrices estimation, and proved
various asymptotic results, while Rothman, Levina and Zhu (2008) generalizes these results to a class of shrinkage operators which includes many commonly used penalty functions.
Wagaman and Levina (2007) developed an
algorithm for finding a meaningful ordering of variables using a
manifold projection technique called the Isomap, so that existing
method like banding can be applied.

The rest of the paper is organized as follows. In section
\ref{sect:BlockPenalizationFramework}, we introduce the model for
block penalization, and the motivation behind. A notion of sign-consistency, we name it block
sign-consistency, is introduced. Together with asymptotic normality,
we call it the oracle property of the resulting one-step estimator. An initial estimator needed for the one-step
estimator, with
the block zero-consistency concept, is introduced in section \ref{subsect:initialestimator}.
A practical algorithm is
discussed, with simulations and real data analysis in
section \ref{sect:Simulations}. Theorems \ref{thm:oracleproperty}(i) and \ref{thm:precisionmatrixconsistency} are proved in the Appendix, whereas Theorems \ref{thm:oracleproperty}(ii) and \ref{thm:initialestimatorBZC} are proved in the Supplement.

\section{Block Penalization Framework}\label{sect:BlockPenalizationFramework}
\setcounter{equation}{0}

\subsection{Motivation}\label{subsect:motivation}
\begin{figure}[htbp]
\centerline{\psfig{figure=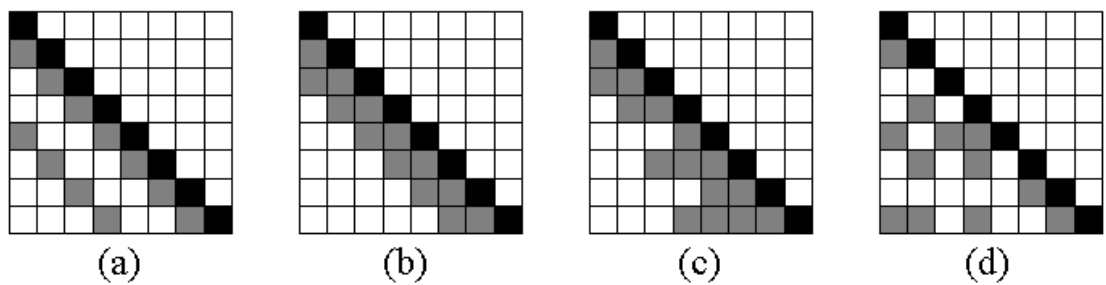,width=6in}}
\begin{singlespace}
\caption{\emph{Pattern of zeros in the resulting estimator for
$\T$ using (a)Block Penalization; (b)Banding; (c)Nested LASSO;
(d)LASSO}}\label{figure1}
\end{singlespace}
\end{figure}

For data with a natural ordering of the variables, e.g.
longitudinal data, or data with a metric equipped like spatial
data with Euclidean distance, if data points are remote in time or
space, they are likely to have weak or no correlation. Then $\T$ in equation (\ref{eqn:MCD}), and thus $\bOmega$,
are banded. Banding and nested LASSO mentioned in
section \ref{sect:introduction} are based on this observation for
obtaining a banded structure of the Cholesky factor $\T$. See
Figure \ref{figure1}(b) for a picture of a banded Cholesky factor.

Also, for variables within a close neighborhood,
the dependence structure should be similar. Equation
(\ref{eqn:MCD}) then says
that coefficients on an off-diagonal band of the Cholesky factor $\T$
are close to neighboring coefficients (see also Wu and Pourahmadi
(2003)). This means that we can improve our estimation if we can
efficiently use neighborhood information (along an off-diagonal band of
$\T$) to estimate the values of individual coefficients.

With these insights, we are motivated to use the block
penalization method. In the context of wavelet coefficients
estimation, Cai (1999) introduced a James-Stein shrinkage rule
over a block of coefficients, whereas Antoniadis and Fan (2001,
page 966) were the first to point out that such method can be regarded as a
special kind of penalized likelihood which penalizes on the $L_2$
norm of a group of coefficients, and introduced a separable
block-penalized least squares for simple solutions. Both papers
argue that block thresholding helps pull information from
neighboring empirical wavelet coefficients, thus increasing the
information available for estimating coefficients within a block.
Yuan and Lin (2006) introduced the same method, which they called
the group LASSO, to select grouped variables (factors) in
multi-factor ANOVA and compare grouped version of LARS and LASSO.
Zhou, Rocha and Yu (2007) further introduced a penalty called the
Composite Absolute Penalty (CAP) to introduce grouping and a
hierarchy at the same time for the estimated parameters in a
linear model.

Block penalization allows for a flexible banded structure in $\T$
since zero off-diagonal bands can precede the non-zero ones.
This is an advantage over banding of Bickel and Levina (2008) and
nested LASSO of Levina \emph{et al.} (2007) as discussed in
section \ref{sect:introduction}. Moreover, the block
sign-consistency property in Theorem \ref{thm:oracleproperty}(i)
implies a banded estimated Cholesky factor $\T$ if the truth $\T_0$ is banded. See Figure \ref{figure1}
for a demonstration.

\vspace{36pt}

\subsection{Block penalization}\label{subsect:model}
As pointed out
in Levina \emph{et al.} (2007), the MCD in (\ref{eqn:MCD}) does
not require the normality assumption of the data, and they introduce a least
squares version for their penalization. We also use such an
approach, and define
\begin{equation}\label{eqn:LS}
L_n(\bphi_n) = \sum_{i=1}^n \sum_{j=2}^{p_n} (y_{ij} -
\y_{i[j]}^T\bphi_{j[j]})^2,
\end{equation}
with $\y_{i[j]} = (y_{i1}, \cdots, y_{i,j-1})^T$, $\bphi_n =
(\bphi_{2[2]}^T,\cdots,\bphi_{p_n[p_n]}^T)^T$, and $\bphi_{j[j]} =
(\phi_{j,1},\cdots,\phi_{j,j-1})^T$.

When $p_{\lambda_n}(\cdot)$ is singular at the origin,
the term-by-term penalty $\sum_{i=2}^{p_n}\sum_{j=1}^{i-1}
p_{\lambda_n}(|\phi_{i,j}|)$ has its singularities located at each
$\phi_{i,j} = 0$, and the block penalty
\begin{equation}\label{eqn:blockpenalty}
J(\bphi_n) = \sum_{j=1}^{p_n-1} p_{\lambda_{nj}}(\| \bl_j \|),
\end{equation}
has its singularities located at $\bl_j = \0$ for $j=1,\cdots,p_n-1$,
where $\lambda_{nj} = \lambda_n(p_n - j)^{1/2}$, $\bl_j =
(\phi_{j+1,1}, \phi_{j+2,2}, \cdots, \phi_{p_n, p_n-j})^T$ is the
$j^{th}$ off-diagonal band of the Cholesky factor $\T$ in
(\ref{eqn:MCD}), and $\|\cdot\|$ is the $L_2$ vector norm.
Hence this block penalty either kills off a whole off-diagonal band
$\bl_j$ or keeps it entirely (see also Antoniadis and Fan (2001)).
%Knowing that $\T$ and $\bOmega$ can be banded for
%longitudinal data, this block penalty takes advantage over
%term-by-term penalty and is shown to perform better than the
%term-by-term LASSO empirically in section \ref{sect:Simulations}.

Combining (\ref{eqn:LS}) and (\ref{eqn:blockpenalty}) is the
block-penalized least squares
\begin{equation}\label{eqn:blockpenalizedLS}
Q_n(\bphin) = L_n(\bphi_n) + nJ(\bphi_n).
\end{equation}
We will use the SCAD penalty function for $p_\lambda(\cdot)$ in
(\ref{eqn:blockpenalty}), defined through its derivative
\begin{equation}\label{eqn:SCAD}
p_\lambda^\prime(\theta) = \lambda\1_{\{\theta \leq \lambda\}} +
(a\lambda - \theta)_+\1_{\{\theta > \lambda\}}.
\end{equation}
SCAD penalty is an unbiased penalty function which has theoretical
advantages over $L_1$-penalty (LASSO). See Lam and Fan (2007) for more details.
In fact, in Fan, Feng and Wu (2007), the SCAD-penalized estimate of a
graphical model is substantially sparser than the $L_1$-penalized
one, which has spuriously large number of edges, partially due to the bias
induced by $L_1$-penalty and hence requiring a smaller $\lambda$ that induces spurious edges. With $\bphinhat$,
we estimate $\D$ in (\ref{eqn:MCD}) by
\begin{equation}\label{eqn:Destimator}
\hat{\sigma}_1^2 = n^{-1}\sum_{i=1}^n y_{i1}^2, \;\;\;
\hat{\sigma}_j^2 = n^{-1}\sum_{i=1}^n (y_{ij} - \y_{i[j]}^T
\boldsymbol{\phihat}_{j[j]})^2,\;\;\; j=2,3,\cdots,p_n.
\end{equation}

\subsection{Linearizing the SCAD penalty}\label{subsect:linearizingSCAD}
Minimizing $Q_n(\bphi_n)$ in (\ref{eqn:blockpenalizedLS}) poses
some challenges. Firstly, $Q_n(\bphi_n)$
is not separable, which makes our problem
computationally challenging. Secondly, the SCAD penalty
complicates the computations as there are no easy
simplifications of the problem like equation (5) in Antoniadis and
Fan (2001, page 966).

Zou and Li (2007) showed that linearizing the SCAD penalty leads
to efficient algorithms like the LARS to be applicable, and that
sparseness, unbiasedness and continuity of the estimators continue
to hold (see Fan and Li (2001)). Following their idea, we
linearize each $p_{\lambda_{nj}}(\| \bl_j \|)$ in
(\ref{eqn:blockpenalty}) at an initial value $\| \blj0 \|$ so that
minimizing (\ref{eqn:blockpenalizedLS}) is equivalent to
minimizing, for $k=0$,
\begin{equation}\label{eqn:linearizedBPloglik}
Q_n^{(k)}(\bphi_n) = \sum_{i=1}^n \sum_{j=2}^{p_n} (y_{ij} -
\y_{i[j]}^T\bphi_{j[j]})^2 + n\sum_{j=1}^{p_n-1}
p_{\lambda_{nj}}^\prime(\| \bl_j^{(k)} \|) \|\bl_j\|,
\end{equation}
where we denote the resulting estimate by $\bphi_n^{(k+1)}$.
Parallel to Theorem 1 and Proposition 1 of Zou and Li (2007), we
state the following theorem concerning convergence in iterating
(\ref{eqn:linearizedBPloglik}) starting from $k=0$.

\begin{theorem}\label{thm:convergence}
For $k = 0,1,2,\cdots$, the ascent property holds for $Q_n$ w.r.t.
$\{ \bphi_n^{(k)} \}$, i.e.
$$ Q_n(\bphi_n^{(k+1)}) \geq Q_n(\bphi_n^{(k)}). $$
Furthermore, let $\bphi_n^{(k+1)} = M(\bphi_n^{(k)})$, so that $M$
is the map carrying $\bphi_n^{(k)}$ to $\bphi_n^{(k+1)}$. If
$Q_n(\bphi_n) = Q_n(M(\bphi_n))$ only for stationary points of
$Q_n$ and if $\bphi_n^*$ is a limit point of the sequence $\{
\bphi_n^{(k)} \}$, then $ \bphi_n^* $ is a stationary point $Q_n$.
\end{theorem}

This convergence result follows from more general convergence
results for MM (minorize-maximize) algorithms. Hence starting from
an initial value $\bphi_n^{(0)}$, we are able to iterate
(\ref{eqn:linearizedBPloglik}) to find a stationary point of
$Q_n$. Note that even starting from the most primitive initial value $\bphi_{j[j]} = \0$, the first step gives a
group LASSO estimator since $p_{\lambda_{nj}}^\prime(0) = \lambda_{nj} = \lambda_n(p_n-j)^{1/2}$. Hence the second step gives a biased reduced estimator of LASSO, as $p_{\lambda_{nj}}^\prime(\| \bl_j^{(k)} \|) = 0$
for $\| \bl_j^{(k)} \| > a \lambda_{nj}$.
In section \ref{subsect:initialestimator} we show how to find a good initial estimator which is theoretically sound,
and iterating until convergence is not always needed.

%Actually (\ref{eqn:linearizedBPloglik}) can be considered a
%weighted block-penalized least squares, with weights $w_{nj}^k =
%np_{\lambda_{nj}}^\prime(\| \bl_j^{(k)} \|) / \lambda_n$. This can be compared
%directly with the adaptive LASSO in Zou (2006), where their
%weights are inversely proportional to the magnitude of initial
%estimators.

\subsection{One-Step Estimator for $\bphi_n$}\label{subsect:onestepestimator}
We now develop a one-step estimator to reduce the computational burden
and prove that such an estimator enjoys the oracle property in
Theorem \ref{thm:oracleproperty}. The performance of this one-step
estimator depends on the initial estimator $\bphin^{(0)}$. Define,
for $\bl_{j0}$ denoting the true value of $\bl_j$ in $\T$,
$$ J_{n0} = \{j : \bl_{j0} = \0  \}, \;\;\; J_{n1} = \{j : \bl_{j0} \neq \0  \}. $$

\setcounter{theorem}{0}
\begin{definition}\label{def:blockzeroconsistency}
An initial estimator $\bphi_n^{(0)}$ is called block
zero-consistent if there exists $\gamma_n = O(1)$ such that (a) $P
\big( \max_{j \in J_{n0}} \| \bl_j^{(0)} \| / (p_n - j)^{1/2} \geq
\gamma_n \big) \rightarrow 0$ as $n \rightarrow \infty$, and (b)
for the same $\gamma_n$, $ P \big( \min_{j \in J_{n1}} \| \bl_j^{(0)} \| / (p_n - j)^{1/2} \geq \gamma_n \big) \rightarrow 1.$
\end{definition}

This definition is similar to the idea of zero-consistency
introduced in Huang, Ma and Zhang (2006), but we now define it at the
block level, which concerns the average magnitude of each element
in the off-diagonal $\bl_j^{(0)}$. With this, we present the main
theorem of this section, the oracle property for the
one-step estimator.

\begin{theorem}\label{thm:oracleproperty}
Assume regularity conditions (A) - (E) in the Appendix, and the Cholesky factor $\T_0$ of the true precision matrix $\bOmega_0$ has
$k_n < n$ non-zero off-diagonal bands. If the initial estimator $\bphi_n^{(0)}$
for $Q_n^{(0)}$ in (\ref{eqn:linearizedBPloglik}) is block
zero-consistent, then the resulting estimator $\bphinhat$ by
minimizing (\ref{eqn:linearizedBPloglik}) satisfies the following:
\begin{itemize}
\item[(i)](Block sign-consistency) $ P(A \cap B) \rightarrow 1 $,
where $A = \{ \boldsymbol{\hat{\ell}}_j = \0 \text{\emph{ for all }} j \in
J_{n0} \},$ and $ B = \{ \text{\emph{sgn}}(\hat{\phi}_{j+k,k}) =
\text{\emph{sgn}}(\phi_{j+k,k}^0) \text{\emph{ for all }} j \in
J_{n1}, \; k \text{\emph{ so that }} \phi_{j+k, k}^0 \neq 0 \}. $

\item[(ii)](Asymptotic normality) Let $\bphi_{n1}$ be the vector
of elements of $\bphi_n$ corresponding to its non-zero off-diagonals. Then
for a vector $\balpha_n$ of the same size as $\boldsymbol{\phihat}_{n1}$ so
that $\balpha_n$ has at most $k_n$ non-zero elements and $\| \balpha_n \| = 1 $, if $k_n^4 (\log^2(k_n+1))^{4/d}/n = o(1)$, we have
$$ n^{1/2} (\balpha_n^T \H_n \balpha_n)^{-1/2} \balpha_n^T( \boldsymbol{\phihat}_{n1} - \bphi_{n1}^0 ) \dcv
N(0, 1),$$
\end{itemize}
where $\H_n$ is block diagonal with $p_n-1$ blocks. Its $(j-1)$-th
block is $\sigma_{j0}^2 \Sigma_{j11}^{-1}$, and $\Sigma_{j11} = E(\y_{i[j]}(1)\y_{i[j]}(1)^T) $, where
$\y_{i[j]}(1)$ contains the elements of $\y_{i[j]}$ corresponding
to the non-zero off-diagonals' elements of $\bphi_{j[j]}^0$.
\end{theorem}

From this theorem and regularity condition (C) in the Appendix, the size $p_n$ of the covariance matrix can be larger than $n$. In particular, if $k_n$ is finite, the oracle property still holds when $\log p_n = o(n)$.
This is useful for many applications with $p_n
> n$, when the sample covariance matrix becomes singular, whereas Theorem \ref{thm:precisionmatrixconsistency} shows that as long as the Cholesky
factor is sparse enough, we can get an optimal estimator of the precision matrix via penalization.

\begin{theorem}\label{thm:precisionmatrixconsistency}
Let
$\hat{\T}$ be the one-step estimator as in Theorem
\ref{thm:oracleproperty}, and $\hat{\D}$ be diagonal with elements
$\hat{\sigma}_{j}^2$ as defined in (\ref{eqn:Destimator}), so that
\mbox{$\hat{\bOmega} = \hat{\T}^T\hat{\D}^{-1}\hat{\T}$}. Then under regularity conditions (A) - (E) in the Appendix, with
$\bOmega_0$ denoting the true precision matrix,
\begin{align*}
\| \hat{\bOmega} - \bOmega_0 \|_{\infty} &= O_P((k_n + 1)^{3/2}(\log p_n/n)^{1/2}), \\
\| \hat{\bOmega} - \bOmega_0 \| &= O_P((k_n + 1)^{5/2} (\log p_n/n)^{1/2}),
\end{align*}
where $ \| M \|_{\infty} = \max_{i,j}|m_{i,j}|$, and $ \| M \| =
\lambda_{\max}^{1/2}(M^TM)$.
\end{theorem}

We will demonstrate related numerical results in section \ref{sect:Simulations}. From
this theorem, the method of block penalization allows for consistent precision matrix
estimation as long as the cost of dimensionality $\log p_n$ satisfies $(k_n+1)^5\log p_n/n = o(1)$.
In particular, if $k_n$ is finite, we only need $\log p_n/n = o(1)$ for consistent estimation. On the other hand, provided the cost of dimensionality is not too large (e.g. $ p_n = n^a$ for some $a > 0$, so $\log p_n = a \log n$ and is negligible), we need
$ k_n = o(n^{1/3}) $ for element-wise consistency.
%The rate is faster than the rate
%at which the number of true non-zero individual coefficients in
%the adaptive LASSO model in Huang, Ma and Zhang (2006) is allowed
%to grow. Moreover, $k_n$ represents the number of non-zero off-diagonal bands
%here, and hence in effect the method of block penalization allows detection
%for much larger number of non-zero coefficients $\phi_{i,j}$ in
%the case of covariance estimation.

\subsection{Block zero-consistent initial estimator}\label{subsect:initialestimator}
We need a block zero-consistent initial estimator for finding an
oracle one-step estimator in the sense of Theorem
\ref{thm:oracleproperty}. The next theorem shows that the OLS
estimator $\tilde{\T}$, where the sample covariance matrix is $\S
= \tilde{\T}^{-1} \tilde{\D} (\tilde{\T}^{-1})^T$ using the MCD in
(\ref{eqn:MCD}), is block zero-consistent when $p_n/n \rightarrow \text{const.} < 1$. When $p_n
>n$, $\S$ is singular and $\tilde{\T}$ is not defined uniquely.
Since we envisage a banded true Cholesky factor $\T_0$ with most non-zero off-diagonals close to
the diagonal, we define $\tilde{\T}$ by considering the least
square estimators of the regression
\begin{equation}\label{eqn:initialestimator}
y_i = \sum_{ j=c_{ni} }^{i-1} \phi_{i,j}
y_j + \epsilon_i,
\end{equation}
where $c_{ni} = \max \{\lfloor i-\gamma n \rfloor, 1 \}$ with some constant $0 < \gamma < 1$ controlling the number of $y_j$'s on which  $y_i$ regresses. The rest of the $\phi_{i,j}$'s are set to zero, recalling that even starting from the most primitive initial value $\bphi_{j[j]} = \0$, the one-step estimator is a
group LASSO estimator since $p_{\lambda_{nj}}^\prime(0) = \lambda_{nj} = \lambda_n(p_n-j)^{1/2}$.
%Then regress $y_i$ on the next $\lfloor \gamma n \rfloor$ $y_j$'s, etc until all the $\tilde{\phi}_{i,j}$'s are obtained.

\begin{theorem}\label{thm:initialestimatorBZC}
Assume regularity conditions (A) to (E) in the Appendix. Then the estimator $\tilde{\T}$ obtained through
the above series of regressions is block
zero-consistent, provided all the true non-zero off-diagonal bands of $\T_0$ are within
the first $\lfloor \gamma n \rfloor$ off-diagonal bands from the main diagonal of $\T_0$.
%The two probabilities converge exponentially fast in definition \ref{def:blockzeroconsistency}, in the sense that
%\begin{align*}
%P(\max_{j \in J_{n0}} \|\boldsymbol{\tilde{\ell}}_j\|/(p_n-j)^{1/2} \geq \gamma_n) &\leq \exp(-o(n)),\\
%P(\min_{j \in J_{n1}} \|\boldsymbol{\tilde{\ell}}_j\|/(p_n-j)^{1/2} \geq \gamma_n) &\geq 1-\exp(-o(n)),
%\end{align*}
%provided $0 < \gamma_n < \min_{j \in J_{n1}} \|\bl_{j0}\|(p_n-j)^{-1/2}.$
\end{theorem}

\emph{Remark} : In high dimensional model selection, the condition of ``irrepresentability''
from Zhao and Yu (2006), ``weak partial orthogonality'' from Huang \etal (2006) or the UUP condition
from Cand\`{e}s and Tao (2007) all describe the need of a weak association
between the relevant covariates and the irrelevant ones under the true model, for the estimation
procedures to pick up the correct sparse signals asymptotically.
In our case, with (\ref{eqn:MCD}) as the true model, the association between the variables $y_i$
and $y_{1},\cdots,y_{i-1}$ for $i=2,\cdots,p_n$ is incorporated into the tail assumption of the $y_{ij}$'s,
which is specified in regularity condition (A). This assumption entails that the $|\phi_{i,j}|$'s
for $i$ and $j$ far apart are small, so that the association between the relevant $y_i$'s
(corr. to $\phi_{t,i} \neq 0$) and the irrelevant $y_j$'s (corr. to $\phi_{t,j} = 0$)
in model (\ref{eqn:MCD}) are small.

In practice, for the series of regression described, we can continue to regress $y_i$ on the next $\lfloor \gamma n \rfloor$ $y_j$'s etc until all the $\tilde{\phi}_{i,j}$'s are obtained. We adapt this initial estimator in the numerical studies in section \ref{sect:Simulations}.

Also in practice, the rate at which $\max_{j \in J_{n0}} \| \bl_j^{(0)}
\| / (p_n - j)^{1/2} $ converges to zero in probability in
definition \ref{def:blockzeroconsistency} may not be fast enough
for the OLS estimators. One way to improve the quality of the OLS
estimators is to smooth along the off-diagonals of $\tilde{\T}$.
For instance, Wu and Pourahmadi (2003) smoothed along off-diagonals
of the OLS estimator $\tilde{\T}$ to reduce estimation errors.
This amounts to assuming that the coefficients $\phi_{i,i-j} =
f_{j,p_n}(i/p_n)$, where $f_{j,p_n}(\cdot)$ is a smooth function defined on $[0,1]$.
We then calculate the smoothed coefficients
$$ \bar{\phi}_{j+k,k} = \sum_{r=1}^{p_n-j} w_j(r+j, k+j) \tilde{ \phi}_{j+r,r},  $$
where the weights $w_j(r+j, k+j)$ depends on the smoothing method. We use
local polynomial smoothing with bandwidth $h \rightarrow \infty$ with $h/p_n \rightarrow 0$,
so that $ \var(\bar{\phi}_{j+k,k}) = O(n^{-1}h^{-1}) $
(See Wu and Pourahmadi (2003) and Fan and Zhang (2000) for more details.).

%\begin{theorem}\label{thm:smoothingT}
%Assume regularity condition (A) in section \ref{sect:Proofs}. For
%the OLS estimator $\tilde{\T}$, and $\nu$ a constant, we have
%$$ P \Big( \max_{j \in J_{n0}} \| \boldsymbol{\tilde{\ell}}_j
%\| / (p_n - j)^{1/2} > \epsilon \Big) = O(p_n \exp(-\nu n
%\epsilon^2)).
%$$
%For the local polynomial smoothed OLS estimator $\bar{\T}$ with
%bandwidth $h$, so that $h \rightarrow \infty$ and $h/p_n
%\rightarrow 0$, we have
%$$ P \Big( \max_{j \in J_{n0}} \| \bar{\bl}_j
%\| / (p_n - j)^{1/2} > \epsilon \Big) = O(q(p_n) \exp(-\nu nh
%\epsilon^2)),
%$$
%where $q(p_n)$ is a polynomial in $p_n$ of fixed order.
%\end{theorem}

%The polynomial $q(\cdot)$ in the theorem is of low order, so that
%the rate $q(p_n)\exp(-\nu nh \epsilon^2)$ goes to zero much faster
%than $p_n\exp(-\nu n \epsilon)$. The performances of both one-step
%estimators resulting from the OLS estimator and the smoothed OLS
%estimator respectively as initial values are compared in section
%\ref{sect:Simulations}.

\subsection{Algorithm for practical implementation}\label{subsect:algorithm}

Yuan and Lin (2006) proposed a group LASSO algorithm to solve
problems similar to (\ref{eqn:linearizedBPloglik}). However, when
$p_n$ is large, the algorithm is computationally very expensive.
Instead, we adapt an idea from Kim, Kim and Kim (2006) and use a
gradient projection method to solve for the one-step estimator,
which is computationally much less demanding. Since minimizing
(\ref{eqn:linearizedBPloglik}) can be considered as a weighted
block-penalized least squares problem with weights $w_{nj}^k =
np_{\lambda_{nj}}^\prime(\| \bl_j^{(k)} \|) / \lambda_n$, it can
be formulated as:
\begin{equation}\label{eqn:constraintminimization}
\text{minimizing } L_n(\bphin) \text{ subject to } \sum_{j=1}^{s_n}w_{nj}^k \| \bl_j \| \leq M
\end{equation}
for some $M \geq 0$. Since the further off-diagonal bands of $\tilde{\T}$ are too short, in
practice we stack them together until it is of length of order
$ p_n $. We then treat it as one block in the above dual-like problem, and denote by $s_n$ the number of off-diagonals in $\tilde{\T}$ after stacking.
%Note that $w_{nj}^k$ depends on $\lambda_n$,
%so that there are two tuning parameters $\lambda_n$ and $M$ to be
%determined. However, for a suitable choice of $\lambda_n$ and that
%$\bl_j = \bl_{j0}$, we either have $w_{nj}^k = 0$ or $\bl_{j0} =
%\0$. Thus the value of $\sum_{j=1}^{s_n}w_{nj}^k \| \bl_{j0} \|$
%is always zero. In view of this, the oracle choice of $M$ is
%actually zero. Hence in fact, we do not need to choose the value
%of $M$, which is ideally zero. This contrasts with the need to
%choose a value for $M$ in the gradient projection algorithm used
%in Kim \etal (2006). We will come to a GCV-type criterion in the
%next section to select $\lambda_n$.

Assume for now that all the tuning parameters are known. Starting from an initial value $\bphi_n^{(0)}$ and
$t=1$, the gradient projection method involves computing the
gradient $\nabla L_n(\bphi_n^{(t-1)})$ and defining $\b =
\bphi_n^{(t-1)} - s \nabla L_n(\bphi_n^{(t-1)}) $, where $s$ is the stepsize of iterations to be found in the next section. Denote by $\b_{(j)}$
the $j$th block of $\b$, with blocks formed according to the
off-diagonals $\bl_j$ of $\T$, $ j = 1,\cdots,s_n $. Then the main
step of the algorithm is to solve
$$ \bphi_n^{t} = \text{argmin}_{\bphin \in \mathcal{B}} \| \b - \bphin \|^2, \;\; \text{ with } \;\;
\mathcal{B} = \Big \{ \sum_{j=1}^{s_n} w_{nj}^k \| \bl_j \| \leq M \Big \}, $$
which is called the projection step. It can be easily reformulated as solving
\begin{equation}\label{eqn:projection}
\min_{M_j} \sum_{j=1}^{s_n} (\| \b_{(j)}\| - M_j )^2 \text{
subject to } \sum_{j=1}^{s_n} w_{nj}^k M_j \leq M, \;\; M_j \geq
0,
\end{equation}
where then $\bl_j^t = M_j \b_{(j)}/\| \b_{(j)} \|$, and we iterate the above until convergence.
Standard LARS or LASSO packages can solve (\ref{eqn:projection}) easily, but we adapt a projection algorithm by Kim \etal (2006) which can solve the above even faster. In solving (\ref{eqn:projection}), we are
essentially projecting $( \| \b_{(1)} \|, \cdots, \| \b_{(s_n)} \| )$ onto the hyperplane
$\sum_{j=1}^{s_n} w_{nj}^k M_j = M$ with $M_j \geq 0$. The key observation is that
if such projection has non-positive values on some $M_j$'s, then the
solution to (\ref{eqn:projection}) should have those $M_j$'s exactly equal zero.
Hence we can then recalculate the projection onto the reduced hyperplane until no more
negative values occur in the projection, and it is easy to see that at most $s_n$ such iterations
are needed to solve (\ref{eqn:projection}). In detail, we start at $\tau = \{1,\cdots,s_n \}$,
and calculate the projection
\begin{equation}\label{eqn:projectedcoordinate}
M_j = \1_{ \{ j \in \tau \} } \Big[ \| \b_{(j)} \| + \Big( M -
\sum_{r \in \tau} w_{nr}^k \| \b_{(r)} \| \Big)
 w_{nj}^k / \sum_{r \in \tau} (w_{nr}^k)^2 \Big]
\end{equation}
for $j = 1,\cdots, s_n$. We then update $\tau = \{ j:  M_j > 0 \}$ and calculate the above projection again
until $M_j \geq 0$ for all $j$.

\subsection{Choice of tuning parameters}\label{subsect:tunepara}
There are three tuning parameters introduced in the previous
section, namely $\lambda_n$, $M$ and $s$.
The
small number $s$ is a parameter for the gradient projection
algorithm and it is required that $s < 2/L$, where $L$ is the
Lipchitz constant of the gradient of $L_n(\bphin)$. It can be easily
shown that $L = 2\lambda_{\max}^{1/2}(S_Y^2)$, where $S_Y =
\diag(\sum_{i=1}^n \y_{i[2]}\y_{i[2]}^T, \cdots,
\sum_{i=1}^n\y_{i[p_n]}\y_{i[p_n]}^T)$, so that $ s <
\lambda_{\max}^{-1/2}(S_Y^2) $.

For the choice of $M$, note that for a suitable $\lambda_n$ and that
$\bl_j = \bl_{j0}$ in (\ref{eqn:constraintminimization}),
we either have $w_{nj}^k = 0$ or $\bl_{j0} = \0$.
Thus, the value of $\sum_{j=1}^{s_n}w_{nj}^k \| \bl_{j0} \|$
is always zero. In view of this, the oracle choice of $M$ is
actually zero. We adapt this choice in the numerical studies in section \ref{sect:Simulations}.

For the choice of $\lambda_n$, we use a GCV criterion similar to
the one used by Kim \etal (2006). We find
$\tilde{\T}$ as defined in section \ref{subsect:initialestimator}, and
smooth the off-diagonal bands of $\tilde{\T}$ to form $\bar{\T}$.
Define $\W_j = \diag(w_{ns_n}^k / \| \bar{\bl}_{s_n} \|
\1_{j-s_n}^T, w_{n(c_{nj}-1)}^k / \| \bar{\bl}_{c_{nj}-1}
\|,\cdots, w_{n2}^k / \| \bar{\bl}_2 \|, w_{n1}^k / \| \bar{\bl}_1
\| )$ and $\X_j = (\y_{1[j]}, \y_{2[j]}, \cdots, \y_{n[j]})^T $, where $\1_m$ denote the column vector of ones of length $m$.
%minimizing
%(\ref{eqn:linearizedBPloglik}) can be seen as minimizing $
%L_n(\bphin) + \lambda_n \sum_{j=1}^{s_n}w_{nj}^k \| \bl_j \|^2 /
%\| \bl_j \|   $, which can then be approximated by, assuming $\|
%\bar{\bl_j} \| \neq 0$,
%\begin{equation}\label{eqn:ridgeapprox}
%L_n(\bphin) + \lambda_n \sum_{j=1}^{s_n}w_{nj}^k \| \bl_j \|^2 / \| \bar{\bl}_j \|.
%\end{equation}
%Minimizing this is a ridge regression problem which can be further
%decomposed into $p_n - 1$ independent minimization problems as in
%Huang \etal (2006): For $j=2,\cdots,p_n$, denote $ c_{nj} =
%\min(s_n, j) $. We minimize, w.r.t. $\bphi_{j[j]}$,
%\begin{equation}\label{eqn:ridgeapproxcomponent}
%\sum_{i=1}^n (y_{ij} - \y_{i[j]}^T\bphi_{j[j]})^2 + \lambda_n
%\bphi_{j[j]}^T \W_j \bphi_{j[j]},
%\end{equation}
%where $\W_j = \diag(w_{ns_n}^k / \| \bar{\bl}_{s_n} \|
%\1_{j-s_n}^T, w_{n(c_{nj}-1)}^k / \| \bar{\bl}_{c_{nj}-1}
%\|,\cdots, w_{n2}^k / \| \bar{\bl}_2 \|, w_{n1}^k / \| \bar{\bl}_1
%\| )$, and $\1_m$ denote the column vector of ones of length $m$.
%The solution is
%\begin{equation}\label{eqn:ridgeapproxsol}
%\bphi_{j[j]}^{\text{ridge}} = (\X_j^T \X_j + \lambda_n
%\W_j)^{-1}\X_j^T \y_j,
%\end{equation}
%where $\X_j = (\y_{1[j]}, \y_{2[j]}, \cdots, \y_{n[j]})^T $ and $
%\y_j = (y_{1j}, y_{2j}, \cdots, y_{nj})^T $.
The GCV-type
criterion is to minimize
\begin{equation}\label{eqn:GCV}
\text{GCV}(\lambda_n) = \sum_{j=2}^{p_n} \frac{n \sum_{i=1}^n
(y_{ij} - \y_{i[j]}^T \bar{\bphi}_{j[j]})^2}{(n - \text{tr}[\X_j
(\X_j^T \X_j + \lambda_n \W_j)^{-1}\X_j^T ])^2},
\end{equation}
where $\tr(\cdot)$ denotes the trace of a square matrix. In practice we calculate GCV($\lambda_n$) on a grid of values of $\lambda_n$ and find the one that minimizes GCV($\lambda_n$) as the solution.

\section{Simulations and Data Analysis}\label{sect:Simulations}
\setcounter{equation}{0}

In this section, we compare the performance of
block penalization (BP) to other regularization methods, in particular banding of Bickel and Levina (2008)
and LASSO of Huang \etal (2006).

For measuring performance, the Kullback-Leibler loss for a precision matrix is used. It has been used in Levina \etal (2007), defined as
$$ L_{KL}(\bSigma, \hat{\bSigma}) = \tr(\hat{\bSigma}^{-1} \bSigma) - \log|\hat{\bSigma}^{-1} \bSigma| - p_n, $$
which is the entropy loss but with the role of covariance matrix and its inverse switched. See Levina \etal (2007) for more details of the loss function.
We also evaluate the operator norm $\| \hat{\bOmega} - \bOmega_0 \|$ for different methods to illustrate the results in Theorem \ref{thm:precisionmatrixconsistency} in our simulation studies. The proportions of correct zeros and non-zeros in the estimators for the Cholesky factors are reported.

\subsection{Simulation analysis}\label{subsect:simulation}
The following three covariance matrices are considered in our simulation studies.
\begin{itemize}
\item[I.]  $\bSigma_1 = 0.8I$.
\item[II.] $\bSigma_2: \phi_{i,i-1} = \phi_{i,i-2} = -0.6, \; \phi_{i,i-4} = \phi_{i,i-6} = -0.4$, $\phi_{i,j} = 0$ otherwise; $\sigma_{j0}^2 = 0.8$.
\item[III.] $\bSigma_3: \phi_{i,j} = 0.5^{i-j}, j < i$; $\sigma_{j0}^2 = 0.1$.
\end{itemize}
The covariance matrix $\bSigma_1$ is a constant multiple of the identity matrix, which is considered by Huang \etal (2006) and
Levina \etal (2007). $\bSigma_2$ is the covariance matrix of an AR(6) process, which has a banded inverse. $\bSigma_3$ is the covariance matrix of an MA(1) process. It is itself tri-diagonal and has a non-sparse inverse. We investigate the performance of BP in such a non-sparse case.

Regularity conditions (B) to (E) are satisfied for the three models by construction. Since all three define stationary
time series models in the sense of (\ref{eqn:MCD}), condition (A) is satisfied from Gaussian to general Weibull-distributed innovations.

We generated $n=100$ observations for each simulation run, and considered $p_n = 50, 100$ and $200$. We used $N=50$ simulation runs throughout.
In order to illustrate theoretical results and test the robustness of the BP method on heavy-tailed data, on top of multivariate normal for the variables, we also consider the multivariate $t_3$ for the variables, which violated condition (A). Tuning parameters for the LASSO and banding are computed using 5-fold CV,
while the parameter $\lambda_n$ for the BP is obtained by minimizing GCV$(\lambda_n)$ in (\ref{eqn:GCV}).
We set the
smoothing parameter $h=0.3$ for local linear smoothing along the off-diagonal bands for demonstration purpose.
The constant $\gamma$ and stacking parameter $s_n$ mentioned in section \ref{subsect:initialestimator} are set at 0.9 and $ p_n - \lceil 2p_n^{1/2} \rceil $ respectively. In fact we have done simulations (not shown) showing that smoothing along off-diagonals for the initial estimator can improve the performance of the one-step estimator. All the results below for the performance of BP are based on such smoothed initial estimators. Also, all subsequent tables show the median of the 50 simulation runs, and the number in the bracket is the $\text{SD}_{\text{mad}}$ which is a robust estimate of the standard deviation, defined by the interquartile range divided by 1.349.

Not shown here, we have carried out comparisons between using GCV-based and 5-fold CV-based tuning parameter $\lambda_n$ for the BP method, and both performed similarly. However, the GCV-based method is much quicker, and hence results of simulations are presented with the GCV-based BP method only.

\begin{table}[t]
  \centering
  \caption{Kullback-Leibler loss for multivariate normal and $t_3$ simulations.}
  \begin{tabular}{ccccc|ccc}\label{table:2}\\
  \hline
   &          & \multicolumn{3}{c|}{Multivariate normal} & \multicolumn{3}{c}{Multivariate $t_3$}\\
   &             $p_n$  & LASSO & Banding & BP &  LASSO & Banding & BP \\
  \hline
 $\bSigma_1$  & 100 & 1.0(.1)	& 1.1(.8) &	1.0(.1)	& 7.7(3.8) &	10.7(9.3)	& 7.8(3.9)\\
             & 200  & 2.1(.2)	& 2.4(3.4)	& 2.1(.2) & 16.4(9.7) &	22.9(18.8)	& 16.4(9.7)\\
  \hline
 $\bSigma_2$ & 100 & 27.2(1.4)	 & 11.1(6.5)	 & 5.6(.5) & 110.7(29.2)	& 57.7(21.1)	& 28.2(10.6)\\
             & 200 & 264.6(39.9)	 & 20.4(12.3)	 & 11.5(.7) & 789.5(132.0) &	101.6(36.0)	& 54.7(14.2)\\
 \hline
 $\bSigma_3$ & 100 & 8.8(.7) & 7.8(9.7)	& 4.3(2.0) & 40.2(7.6)	& 31.8(14.9)	& 19.8(7.9)\\
             & 200 & 19.4(1.5)	& 24.9(83.4)	& 18.1(23.1) & 99.6(23.6)	& 70.3(35.4)	& 56.3(26.0)\\
\hline
\end{tabular}
\end{table}

Table \ref{table:2} shows the Kullback-Leibler loss from various methods for multivariate normal and $t_3$ simulations. We omit the case for $p_n = 50$ to save space, but results are similar to those for higher dimensions. In general the higher the dimension, the larger the loss is for all the methods. On $\bSigma_1$, all methods perform
similarly as expected (sample covariance matrix performs much worse and is not shown). However on $\bSigma_2$, BP performs much better for all $p_n$ considered, especially when multivariate $t_3$ is concerned. The better performance is expected, since BP can eliminate weaker signals that precede stronger ones, but not particularly so for other methods. On $\bSigma_3$, BP performs slightly better on average, particularly
for multivariate $t_3$ simulations. For normal data, LASSO has smaller variability, though.

\begin{table}[t]
  \centering
  \caption{Operator norm of difference $\|\hat{\bOmega} - \bOmega_0\|$ for different methods.}
  \begin{tabular}{ccccc|ccc}\label{table:4}\\
  \hline
 \multicolumn{2}{c}{} & \multicolumn{3}{c|}{Multivariate normal} & \multicolumn{3}{c}{Multivariate $t_3$}\\
   &             $p_n$  & LASSO & Banding & BP & LASSO & Banding & BP \\
  \hline
 $\bSigma_1$ & 100 & .6(.1) & .7(.3) &  .6(.1)  & 1.7(.5)	& 2.0(.8)	 & 1.7(.5)\\
             & 200 & .7(.1) & .8(.4) &	.7(.1)  & 1.8(.6)	& 2.0(.9)	 & 1.8(.5)\\
  \hline
 $\bSigma_2$  & 100 & 5.9(.4)	& 6.2(3.5)  &	 2.5(.4) & 11.3(4.6) & 11.0(6.6)	& 7.2(3.5)\\
             & 200 & 29.1(11.3) &	5.7(3.4)  & 2.6(.4) & 58.1(11.2)	& 12.1(5.7)	& 7.7(2.3)\\
 \hline
 $\bSigma_3$ & 100 &14.7(1.6) & 19.0(14.2) & 11.6(1.9) & 40.3(9.1)	& 33.8(13.5) & 28.1(6.6)\\
             & 200 &16.0(1.4) & 27.4(63.7) & 18.4(6.1) & 46.1(6.0)  & 42.2(17.3) & 35.5(11.0)\\
\hline
\end{tabular}
\end{table}

To demonstrate results of Theorem \ref{thm:precisionmatrixconsistency}, the operator norm of difference
$\|\hat{\bOmega} - \bOmega_0\|$ for different methods are summarized in Table \ref{table:4}. Clearly BP performs better in comparison with LASSO and banding on $\bSigma_2$, in both normal and $t_3$ innovations. The performance gap gets larger as $p_n$ increases. For $\bSigma_3$ BP still outperforms the other two methods in general, especially for heavy-tailed data.

%Theorem \ref{thm:precisionmatrixconsistency} states that $\|\hat{\bOmega} - \bOmega_0\|^2 = O_P((k_n+1)^3\log p_n/n)$ for the BP method. With
%$n=100$ and $k_n=4$ for the Cholesky factor of $\bSigma_2$, the values of $((k_n+1)^3\log p_n/n)^{1/2}$ are 2.40 and 2.57 for $p_n = 100$ and 200 respectively, which is remarkably close to the simulated median values for the normal simulations. For $\bSigma_1$, $k_n=0$ and the values are 0.21 and 0.23 respectively for $p_n = 100$ and 200.

\begin{table}[b]
  \centering
  \caption{Correct zeros and non-zeros(\%) in the estimated Cholesky factors for $\bSigma_2$.}
  \begin{tabular}{ccccc|ccc}\label{table:5}\\
  \hline
 \multicolumn{2}{c}{} & \multicolumn{3}{c|}{Multivariate normal} & \multicolumn{3}{c}{Multivariate $t_3$}\\
\hline
   &             $p_n$  & LASSO & Banding & BP & LASSO & Banding & BP \\
  \hline
 Correct & 50  & 60.6(2.3)	& 73.5(20.1)	& 100(0) & 56.5(3.5)	& 89.1(12.3)	& 95.6(14.0)\\
percentage          & 100 &75.3(.9)	& 87.7(12.0)	& 100(0)  & 70.5(2.6)	& 94.4(5.8)	& 100(0)\\
 of zeros         & 200 &73.5(.7)	& 92.9(8.7)	& 100(0) & 72.0(.7)	& 97.3(2.7)	& 100(0)\\
  \hline
Correct & 50  &99.6(.4) &	100(0) &	100(0) & 96.4(1.6) &	71.3(35.0) &	100(0)\\
 percentage            & 100 &99.2(.3) &	100(0) &	100(0) & 95.1(1.8)	& 72.3(33.3)	& 100(0)\\
 of non-zeros            & 200 &99.3(.3) &	100(0) &	100(0) & 97.1(.7)  &	80.5(25.9)	& 100(0)\\
 \hline
\end{tabular}
\end{table}

Finally, to illustrate the ability to capture sparsity, we focus on $\bSigma_2$ and summarize the correct percentages of zeros and non-zeros
estimated in Table \ref{table:5}. BP almost gets all the zeros and non-zeros right in all simulations. The LASSO does poorly in
the correct percentages of zeros. This is due to biases induced by LASSO that require a relatively small $\lambda$, resulting in many spurious non-zero coefficients. The banding method does not work well too. However, note that both banding and BP do better as dimension increases.

\subsection{Real data analysis}\label{subsect:realdata}
We analyze the call center data using the BP method. This set of data is described in detail and analyzed by Shen and Huang (2005), and we thank you for the data courtesy by the authors.

The original data consists of details of every call to a call center of a major northeastern U.S. financial firm in 2002. Removing calls from weekends, holidays, and days when recording equipment was faulty, we obtain data from 239 days. On each of these days, the call center open
from 7am to midnight, so there is a 17-hour period for calls each day. For ease of comparison, following Huang \etal (2006) and
Bickel and Levina (2008), we use the data which is divided into 10-minute intervals, and the number of calls in each interval is denoted by $N_{ij}$, for days $i=1,\cdots, 239$ and interval $j=1,\cdots,102$. The transformation $y_{ij} = (N_{ij} + 1/4)^{1/2}$ is used to make the data closer to normal.

\begin{figure}[b]
\centerline{\psfig{figure=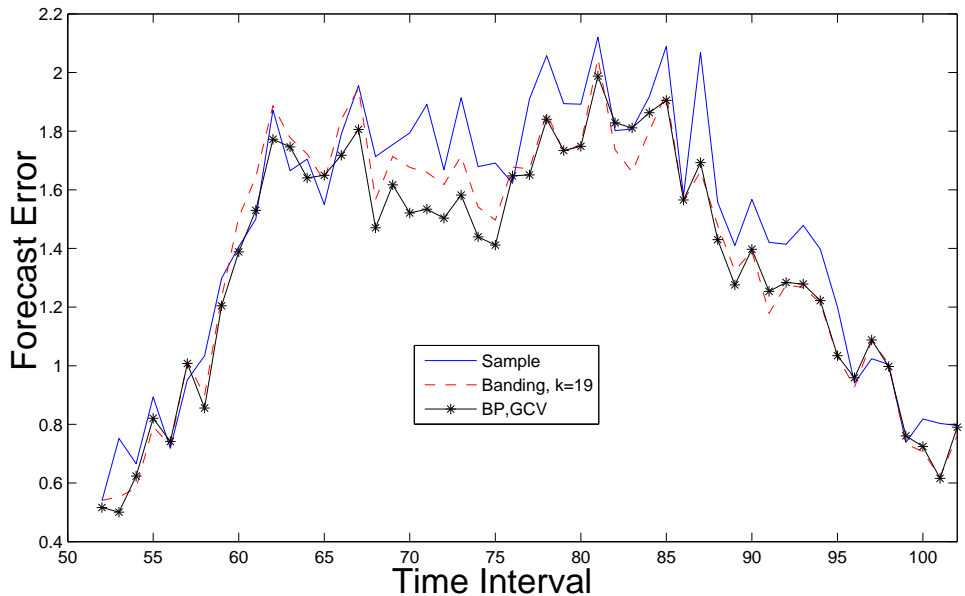,width=6in}}
\begin{singlespace}
\caption{\emph{Mean absolute forecast errors for different estimation methods. Average is taken over 34 days of test data from November to December, 2002.}}\label{figure2}
\end{singlespace}
\end{figure}

The goal is to forecast the counts of arrival calls in the second half of the day from those in the first half of the day.
If we assume $\y_i = (y_{i1},\cdots,y_{i,102})^T \sim N(\bmu, \bSigma)$, partitioning $\y_i$ into $\y_i^{(1)}$ and $\y_i^{(2)}$ where
$\y_i^{(1)} = (y_{i1},\cdots,y_{i,51})^T,  \y_i^{(2)} = (y_{i,52},\cdots,y_{i,102})^T$, and denoting
\begin{equation*}
\bmu = \left(
  \begin{array}{c}
    \bmu_1 \\
    \bmu_2 \\
  \end{array}
\right), \;\;
\bSigma = \left(
   \begin{array}{cc}
     \bSigma_{11} \;\; \bSigma_{12}\\
     \bSigma_{21} \;\; \bSigma_{22}\\
   \end{array}
\right),
\end{equation*}
the best mean square error forecast is then given by the conditional mean
$$ \hat{\y}^{(2)} = E( \y^{(2)} | \y^{(1)} ) = \hat{\bmu}_2 + \hat{\bSigma}_{21} \hat{\bSigma}_{11}^{-1}(\y^{(1)} - \hat{\bmu}_1). $$
This is also the best mean square error linear predictor without normality assumption.

To compare performance of different estimators of $\bSigma$, we divide the data into a training set (Jan. to Oct., 205 days) and a test set (Nov. and Dec., 34 days). We estimate $\hat{\bmu} = \sum_{i=1}^{205} \y_i/205$, and $\hat{\bSigma}$ by sample covariance, banding and BP. For
each time interval $j = 52,\cdots,102$, we consider the mean absolute forecast error
$$ \text{Err}_j = \frac{1}{34} \sum_{i=206}^{239} |\hat{y}_{ij} - y_{ij}|. $$
For BP, we use GCV with $h=0.1$. The number $k=19$ for banding is used in Bickel and Levina (2008). From Figure \ref{figure2}, it is clear that the BP outperforms the other two methods, in particular for the time intervals 66 to 75 corresponding to the mid-afternoon.

\vspace{24pt}
\centerline{\large \bf Appendix: Proof of Theorems \ref{thm:oracleproperty}(i) and \ref{thm:precisionmatrixconsistency}}
\setcounter{equation}{0}
\renewcommand \theequation{A.\arabic{equation}}

We state the following general regularity conditions for the results
in section \ref{sect:BlockPenalizationFramework}.
\begin{itemize}
\item[(A)] The data $\y_i, i=1,2,\cdots,n$ are i.i.d. with mean
zero and variance $\bSigma_0$, a symmetric positive-definite matrix
of size $p_n$. The tail probability of $\y_i$ satisfies, for $ j=1,2,\cdots,p_n $,
$\mbox{$P(|y_{ij}| > x) \leq K\exp(-Cx^d)$, } $
where $d > 0$ and $C$, $K$ are constants. The innovations
$\epsilon_{i2}, \cdots, \epsilon_{ip_n}$ for $i=1,\cdots,n$ in (\ref{eqn:MCD}) are mutually independent zero-mean
r.v.'s and $\var(\epsilon_{ij}) = \sigma_{j0}^2$, having tail probability bounds similar to the
$y_{ij}$'s.

\item[(B)] The variance-covariance matrix $\bSigma_0$ in (A) has
eigenvalues uniformly bounded away from 0 and $\infty$ w.r.t. $n$.
That is, there exists constants $C_1$ and $C_2$ such that
$$ 0 < C_1 < \lambda_{\min}(\bSigma_0) \leq \lambda_{\max}(\bSigma_0) < C_2 < \infty \;\;\; \text{for all } n, $$
where $\lambda_{\min}(\bSigma_0)$  and $\lambda_{\max}(\bSigma_0)$ are the minimum and maximum eigenvalues of $\bSigma_0$ respectively.

\item[(C)] Let $ d_{n1} = \min \{ \phi_{n1j}^0: \phi_{n1j}^0 > 0 \}
$, where $\phi_{n1j}^0$ is the $j$-th element of $\bphi_{n1}^0$ (see Step 2.1 in the proof of Theorem \ref{thm:oracleproperty}(i) for a definition).
Then as $n \rightarrow \infty$,
$$ \frac{k_n \log p_n}{n d_{n1}^2} \rightarrow 0,
\quad \frac{k_n^2 \log p_n}{n \lambda_n} \rightarrow 0, \quad \frac{\log p_n}{n \lambda_n^2}
\rightarrow 0.
$$

\item[(D)] The tuning parameter $\lambda_n$ satisfies
\begin{equation*}
0 < \lambda_n < \min_{j \in J_{n1}} \frac{\| \bl_{j0} \|}{ a(p_n - j)^{1/2}},
\end{equation*}
with $(p_n - j) \rightarrow \infty $ for all $j \in J_{n1}$ as $n
\rightarrow \infty$.

\item[(E)] The values $\sigma_{\epsilon M}^2 = \max_{1 \leq t \leq p_n} \sigma_{t0}^2$ and
$\sigma_{yM}^2 = \max_{1 \leq r \leq p_n} \var(y_{jr})$
are bounded uniformly away from zero and infinity.
\end{itemize}

%Condition (A) allows data with different tails behaviors to be
%considered, which ranges from light tails (d=2 for Gaussian tail to d=1 for exponential tail) to heavy tails where $0 < d < 1$.
%Banded true Cholesky factor $\T_0$ allows
%for the $\epsilon_{ij}$'s and the $y_{ij}$'s to have similar tail
%probability behaviors. We assume $K$ and $C$ to be constants in
%the bound which implicitly constrained the growth of
%$\phi_{i,j}^0$'s in a non-zero off-diagonal of $\T_0$. Also this
%implicitly assumes that the $\epsilon_{ij}$'s and the $y_{ij}$'s
%have second and fourth moments bounded uniformly away from
%zero and infinity as $j$ varies. This can actually be
%relaxed at the expense of more complicated proof and formulas in
%condition (C). Condition (B) bounds uniformly the eigenvalues of
%the true covariance matrix as $n$ grows. Indeed these eigenvalues
%can grow w.r.t. $n$ at certain rates, but we do not consider it
%here in return for shorter proofs. Condition (C) controls the
%amount of non-zero off-diagonals allowed in $\T_0$ and the size of the true
%covariance matrix. Condition (D) allows for an oracle choice of
%$\lambda_n$ when the smallest (in terms of $L_2$ norm) non-zero
%off-diagonal is known, when the penalty function is the SCAD
%penalty as defined in (\ref{eqn:SCAD}). Certainly other choices of penalty functions are allowed
%and parallel condition can be derived. Condition (E)
%is used for all the proofs of the Theorems in the paper, which also constrains
%the growth of the $\phi_{i,j}^0$'s implicitly.

The following lemma is a direct consequence of Theorem 5.11 of Bai and Silverstein (2006).
\begin{lemma}\label{lemma:extremeevalues}
Let $\{\y_i\}_{1 \leq i \leq n}$ be a random sample of $n$ vectors with length $q_n$, each with mean $\0$ and covariance matrix $\bSigma$. In addition, each element of $\y_i$ has finite fourth moment. Then if $q_n/n \rightarrow \ell < 1$, the sample covariance matrix $\S_n = n^{-1}\sum_{i=1}^n \y_i\y_i^T$ satisfies, almost surely,
$$ \lim_{n \rightarrow \infty} \lambda_{\max}(\S_n) \leq \lambda_{\max}(\bSigma) (1 + \sqrt{\ell})^2, \;\;
  \lim_{n \rightarrow \infty} \lambda_{\min}(\S_n) \geq \lambda_{\min}(\bSigma) (1 - \sqrt{\ell})^2. $$
\end{lemma}

\vspace{12pt}
\noindent {\bf Proof of Lemma \ref{lemma:extremeevalues}}.
By Theorem 5.11 of Bai and Silverstein (2006), the matrix $\S_n^* = \bSigma^{-1/2} \S_n \bSigma^{-1/2}$ which is the sample covariance matrix of $\bSigma^{-1/2}\y_i$, has
$$ \lim_{n \rightarrow \infty} \lambda_{\max} (\S_n^*) = (1 + \sqrt{\ell})^2, \;\;
 \lim_{n \rightarrow \infty} \lambda_{\min} (\S_n^*) = (1 - \sqrt{\ell})^2 $$
almost surely. Since $\ell < 1$, this implies that $\S_n^*$ is almost surely invertible. Then by standard arguments,
$$ \lim_{n \rightarrow \infty} \lambda_{\min}(\S_n) = \lim_{n \rightarrow \infty} \lambda_{\min} (\bSigma^{1/2} \S_n^* \bSigma^{1/2}) \geq
\lambda_{\min} (\bSigma) (1-\sqrt{\ell})^2 $$
almost surely. The other inequality is proved similarly. $\square$

\vspace{12pt}
\noindent {\bf Proof of Theorem \ref{thm:oracleproperty}}.
The idea is to prove that the probability of a sufficient condition for block-sign consistency approaches 1 as $n \rightarrow \infty$.
We split the proof into multiple steps and substeps to enhance readability. We prove for the case $k_n \geq 1$ first, with the case $k_n = 0$ put at the end of the proof.

\indent {\bf Step 1.} \; \emph{Sufficient condition for solution to exist.}
An elementwise sufficient condition, derived from the Karush-Kuhn-Tucker (KKT) condition for $ \boldsymbol{\phihat}_n $ to be a solution to minimizing (\ref{eqn:linearizedBPloglik})  (see for example Yuan and Lin (2006) for the full KKT condition),
is
\begin{align}\label{eqn:KKT1}
%\begin{split}
2 \sum_{i=1}^n y_{i,t-j} (y_{it} - \y_{i[t]}^T \boldsymbol{\phihat}_{t[t]}) &= \lambda_n w_{nj}^k \hat{\phi}_{t,t-j}/\| \boldsymbol{\hat{\ell}}_j \|,
 \; \text{ for all } \boldsymbol{\hat{\ell}}_j \neq \0, \\ \label{eqn:KKT2}
\bigg| 2 \sum_{i=1}^n y_{i,t-j} (y_{it} - \y_{i[t]}^T
\boldsymbol{\phihat}_{t[t]})  \bigg|
 &\leq \lambda_n w_{nj}^k (p_n - j)^{-1/2}, \; \text{ for all } \boldsymbol{\hat{\ell}}_j = \0,
%\end{split}
\end{align}
where $ t = j+1, \cdots, p_n$ and $w_{nj}^k = np_{\lambda_{nj}}^\prime(\| \bl_j^{(k)} \|)/\lambda_n$ (see section \ref{subsect:model} for more definitions). We assume WLOG that the $k_n$ non-zero off-diagonals of the true Cholesky factor
$\T_0$ are its first $k_n$ off-diagonals to simplify notations. We also assume no stacking (see section \ref{subsect:algorithm}) of the last off-diagonal bands of $\T$ in solving (\ref{eqn:linearizedBPloglik}); the case of stacked off-diagonals can be treated similarly.

\vspace{12pt}
\indent {\bf Step 2.} \; \emph{Sufficient condition for block sign-consistency.} \: To introduce the sufficient condition for block-sign consistency, we define
$\C_{tjk} = n^{-1} \sum_{i=1}^n \y_{i[t]}(j) \y_{i[t]}(k)^T$ for $j,k = 1,2$, where $\y_{i[t]}(2)$ contains
the elements of $\y_{i[t]}$ corresponding to the zero off-diagonals' elements of $\bphi_{t[t]}^0$, and $\y_{i[t]}(1)$ contains the rest. We also define, for $t=2,\cdots,p_n$,
\begin{align*}
\v_t &= n^{-1/2} \sum_{i=1}^n \epsilon_{it} \y_{i[t]}, \;\;
\epsilon_{it} = y_{it} - \y_{i[t]}^T \bphi_{t[t]}^0, \;\; \W_{nt} =
\diag ( w_{nb_{nt}}^k, \cdots, w_{n2}^k, w_{n1}^k  ), \\
\tilde{\w}_{nt} &=  (\tilde{w}_{n(t-1)}^k, \cdots, \tilde{w}_{n1}^k)^T, \;\;
\s_t = (\hat{\phi}_{t,t-b_{nt}}/ \| \boldsymbol{\hat{\ell}}_{b_{nt}} \|, \cdots, \hat{\phi}_{t,t-2}/ \| \boldsymbol{\hat{\ell}}_2 \|,
\hat{\phi}_{t,t-1}/ \| \boldsymbol{\hat{\ell}}_1 \|)^T,
\end{align*}
where $b_{nt} = \min(t-1, k_n)$, $\tilde{w}_{nj}^k = w_{nj}^k(p_n - j)^{-1/2}$. Also, $\v_t(j), \tilde{\w}_{nt}(j)$ for $j=1,2$ are defined similar to $\y_{i[t]}(j)$;
$\bphi_{t[t]}^0(j)$ and $\boldsymbol{\phihat}_{t[t]}(j)$ for $j=1,2$ are defined similarly also.

For $\bphinhat$ to be block sign-consistent, we need only to show
that equation
(\ref{eqn:KKT1}) is true for $j=1,\cdots,k_n$, equation
(\ref{eqn:KKT2}) is true for $j=k_n+1,\cdots,p_n-1$, and
$|\boldsymbol{\phihat}_{t[t]}(1) - \bphi_{t[t]}^0(1)| < |\bphi_{t[t]}^0(1)|$.
It is sufficient to show that the
following conditions occur with probability going to 1 (this is similar to Zhou and Yu
(2006) Proposition 1; see their paper for more details):
\begin{equation}\label{eqn:suffcondforBSC}
\begin{split}
| \C_{t11}^{-1} \v_t(1) | &< n^{1/2}|\bphi_{t[t]}^0(1)| - \lambda_n n^{-1/2} \C_{t11}^{-1} \W_{nt} \s_t /2, \\
|\C_{r21} \C_{r11}^{-1} \v_r(1)  - \v_r(2) | &\leq \lambda_n n^{-1/2} (\tilde{\w}_{nr}(2) - |\C_{r21} \C_{r11}^{-1} \W_{nr} \s_r|) /2,
\end{split}
\end{equation}
where $t=2,\cdots,p_n$ and $r = k_n+2,\cdots, p_n$. Since the matrix $\C_{t11}$ has size at most $k_n$ and $k_n/n = o(1)$, $\C_{t11}$ is almost surely invertible as $ n \rightarrow \infty $ by Lemma \ref{lemma:extremeevalues} and condition (B). In more compact form, it can be written as
\begin{equation}\label{eqn:suffcondforBSCcompact}
\begin{split}
| \G_{11}^{-1} \z | &< n^{1/2}|\bphi_{n1}^0| - \lambda_n n^{-1/2} \G_{11}^{-1} \W_n \s /2,\\
| \G_{21} \G_{11}^{-1}(2) \z(2) - \tilde{\z}| &\leq \lambda_n n^{-1/2} (\tilde{\w}_n - | \G_{21} \G_{11}^{-1}(2) \W_n(2) \s(2) |) /2,
\end{split}
\end{equation}
where
\begin{align*}
\G_{11} &= \diag(\C_{211},\cdots,\C_{p_n11}), \;\;\;\;\;\;\;\;\;\: \G_{21} = \diag(\C_{(k_n+2)21},\cdots,\C_{p_n21}),\\
\G_{11}(2) &= \diag(\C_{(k_n+2)11}, \cdots, \C_{p_n11}), \;\;\;\;\;\;\; \z = (\v_2(1)^T,\cdots,\v_{p_n}(1)^T)^T,\\
\z(2) &= (\v_{k_n+2}(1)^T,\cdots,\v_{p_n}(1)^T)^T, \:\;\;\;\;\;\;\: \tilde{\z} = (\v_{k_n+2}(2)^T,\cdots,\v_{p_n}(2))^T,\\
\bphi_{n1}^0 &= (\bphi_{2[2]}^{0}(1)^T,\cdots, \bphi_{p_n[p_n]}^{0}(1)^T)^T, \; \W_n = \diag(\W_{n2}, \cdots, \W_{np_n}),\\
\W_n(2) &= \diag(\W_{n(k_n+2)}, \cdots, \W_{np_n}), \;\;\;\;\;\;\; \s = (\s_2^T, \cdots, \s_{p_n}^T)^T,\\
\s(2) &= (\s_{k_n+2}^T, \cdots, \s_{p_n}^T)^T, \;\;\quad\quad\quad\;\;\;\;\;\; \tilde{\w}_n = (\tilde{\w}_{n(k_n+2)}(2)^T, \cdots, \tilde{\w}_{np_n}(2)^T)^T.
\end{align*}

\vspace{12pt}
\indent {\bf Step 3.} \: Denote by $A_n$ and $B_n$ respectively the events that the first and the second conditions of (\ref{eqn:suffcondforBSCcompact}) hold. It is sufficient to show $P(A_n^c) \rightarrow 0$ and $P(B_n^c) \rightarrow 0$ as $n \rightarrow \infty$.

\indent \emph{Step 3.1} \; \emph{Showing $P(A_n^c) \rightarrow 0$.} \:
Define $\boldeta = \G_{11}^{-1} \z$, and $\boldeta_n = \G_{11}^{-1} \z_n$, where $\z_n = (z_{n,j})_{j \geq 1}^T$ with
$z_{n,j} = n^{-1/2} \sum_{i=1}^n y_{ir} \epsilon_{it}\1_{\{ |y_{ir}|,|\epsilon_{it}| \leq a(n) \}}$, a truncated version of
$ z_j = n^{-1/2} \sum_{i=1}^n y_{ir} \epsilon_{it} $ for some $r$, $t$ with $\max(1, t-k_n) \leq r < t$. Denote by $\eta_{n,j}$ the $j$-th element of $\boldeta_n$. In these definitions, $a(n) \rightarrow \infty$ as $n \rightarrow \infty$.

We need the following result, which will be shown in Step 5:
\begin{equation}\label{eqn:expectedetabound}
E(\max_j| \eta_{n,j}|) = O((k_n \log p_n)^{1/2} a^2(n)).
\end{equation}

Since the initial estimator $\bphi_{n}^{(k)}$ in
(\ref{eqn:linearizedBPloglik}) is block zero-consistent, if
$\lambda_n$ is chosen to satisfy condition (D), then $\gamma_n$ in
Definition \ref{def:blockzeroconsistency} can be set to this
$\lambda_n$. It is easy to see that
\begin{equation}\label{eqn:weightforBZC}
P(\tilde{w}_{nj}^k = n, \; \forall j \in J_{n0}) \rightarrow 1,
\;\; P(w_{nj}^k = 0, \; \forall j \in J_{n1}) \rightarrow 1 \; \text{ as } \; n \rightarrow \infty.
\end{equation}

By definition, $\boldeta_{n} - \boldeta \rightarrow \0$ almost surely as $n \rightarrow \infty$. Thus,
$ \1_{\{ \max_j |\eta_{n,j}| \geq n^{1/2} d_{n1} \}} - \1_{\{ \max_j |\eta_{j}| \geq n^{1/2} d_{n1} \}} \rightarrow 0$ almost surely,
implying
\begin{equation}\label{eqn:probconv1}
P(\max_j |\eta_{n,j}| \geq n^{1/2} d_{n1}) - P(\max_j |\eta_{j}| \geq n^{1/2} d_{n1}) \rightarrow 0 \text{ as } n \rightarrow \infty.
\end{equation}
Then by the Markov inequality and (\ref{eqn:expectedetabound}),
\begin{align*}
P\big( \max_j |\eta_{n,j}| \geq n^{1/2} d_{n1} \big) &\leq E\big(\max_j |\eta_{n,j}| \big)/(n^{1/2}d_{n1})\\
 &=O(( k_n \log p_n)^{1/2} a^2(n) / (n^{1/2} d_{n1})) \rightarrow 0,
\end{align*}
by condition (C) and for $a(n)$ chosen to go to infinity slow enough. Hence by (\ref{eqn:probconv1}), we have $P\big( \max_j |\eta_j| \geq n^{1/2} d_{n1} \big) \rightarrow 0$, thus
\begin{align*}
P(A_n^c) &\leq P(A_n^c \cap \{ w_{nj}^k = 0, \; \forall j \in J_{n1} \}) + P( w_{nj}^k > 0, \; \forall j \in J_{n1} ) \\
&\leq P( \max_j |\eta_j| \geq n^{1/2} d_{n1} ) + P( w_{nj}^k > 0, \; \forall j \in J_{n1} ) \rightarrow 0,
\end{align*}
using (\ref{eqn:weightforBZC}) and the fact that
\begin{align*}
A_n^c \cap \{ w_{nj}^k = 0 \; \forall j \in J_{n1} \} &= \{|\G_{11}^{-1} \z| \geq n^{1/2} |\bphi_{n1}^0| \} \subset \big\{ \max_j |\eta_j| \geq n^{1/2} d_{n1} \big\}.
\end{align*}

\indent \emph{Step 3.2} \; \emph{Showing $P(B_n^c) \rightarrow 0$.} \:
Define $\bzeta = \G_{21} \G_{11}^{-1}(2) \z(2)$,
then $\zeta_j = (\C_{t21} \C_{t11}^{-1} \v_t(1))_r$ for some $t$, $r$ with $t \geq k_n+2$.
Also, define $x_{rk} = n^{-1/2} \sum_{i=1}^n y_{ir} y_{ik}$, and $x_{n,rk}$ the truncated version (by $a(n)$) similar to $z_{n,j}$ in Step 3.1.
Then  we can rewrite $\zeta_j = n^{-1/2} \sum_{k} x_{rk} \eta_{k}$, and define
$$ \zeta_{n,j} = n^{-1/2} \sum_{k} x_{n,rk} \eta_{n,k}, $$
for some $r$. The summation involves at most $k_n$ terms.

We need the following results, which will be shown in Step 4 and 6 respectively:
\begin{align}
E(\max_k |z_{n,k}|) &= O((\log p_n)^{1/2} a^2(n)), \label{eqn:expectedzbound} \\
E(\max_j |\zeta_{n,j}|) &= O(k_n^2 \log p_n a^4(n)). \label{eqn:expectedzetabound}
\end{align}

By definition, for all $j$, $ \zeta_{n,j} - \zeta_j \rightarrow 0$ and $ z_{n,j} - z_j \rightarrow 0$ almost surely, implying
\begin{equation}\label{eqn:probconv2}
P(\max_{j,k} |\zeta_{n,j} - z_{n,k}| \geq \lambda_n n^{1/2}/2) - P(\max_{j,k} |\eta_{j} - z_k| \geq \lambda_n n^{1/2}/2) \rightarrow 0 \text{ as } n \rightarrow \infty.
\end{equation}
Then by the Markov inequality, (\ref{eqn:expectedzbound}) and (\ref{eqn:expectedzetabound}),
\begin{align*}
P(\max_{j,k} |\zeta_{n,j} - z_{n,k}| &\geq \lambda_n n^{1/2} /2) \leq 2 \{ E(\max_j |\zeta_{n,j}|) + E(\max_k|z_k|) \} /(\lambda_n n^{1/2})\\
&=O(k_n^2 \log p_n \cdot a^4(n) / (\lambda_n n) + (\log p_n)^{1/2} a^2(n) / (\lambda_n n^{1/2})),
\end{align*}
which goes to 0 by condition (C), for $a(n)$ chosen to go to infinity slow enough. This implies
$P(\max_{j,k} |\zeta_{j} - z_k| \geq \lambda_n n^{1/2} /2) \rightarrow 0$ by (\ref{eqn:probconv2}).

Define $D_n = \{ \tilde{w}_{nj}^k = n \; \forall j \in J_{n0} \}
\cap \{ w_{nj}^k = 0 \; \forall j \in J_{n1} \} $, so that
$P(D_n^c) \rightarrow 0$ by (\ref{eqn:weightforBZC}). Hence using
$B_n^c \cap D_n = \{ |\bzeta - \tilde{\z}| \geq \lambda_n n^{1/2} /2  \}
\subset \big\{ \max_{j,k} |\zeta_j - z_k| \geq \lambda_n n^{1/2} /2 \big\},$
\begin{align*}
P(B_n^c) &\leq P(B_n^c \cap D_n) + P(D_n^c)\\
&\leq P(\max_{j,k} |\zeta_j - z_k| \geq \lambda_n n^{1/2} /2) + P(D_n^c) \rightarrow 0.
\end{align*}

\vspace{12pt}
\indent {\bf Step 4.} \; \emph{Proof of (\ref{eqn:expectedzbound}).} \:
This requires the application of Orlicz norm of a random variable $X$,
which is defined as
$\| X \|_{\psi} = \inf\{ C>0 : E{\psi}(|X|/C) \leq 1 \} $, where $\psi$ is a non-decreasing convex function with $\psi(0) = 0$. We define
 $\psi_a(x) = \exp(x^a) -1$ for $a \geq 1$,
which is non-decreasing and convex with $\psi_a(0) = 0$.
See section 2.2 of van der Vaart and Wellner (2000) (hereafter VW(2000)) for more details.
%We consider bounding various variables by the Orlicz norm in the remainder of Step 3.

We need four more general results on Orlicz norm:

1. By Proposition A.1.6 of VW (2000), for any independent zero-mean r.v.'s $W_i$, define $S_n = \sum_{i=1}^n W_i$, then
\begin{align}
\| S_n \|_{\psi_1} &\leq K_1 \big( E|S_n| + \| \max_{1 \leq i \leq n}|W_i| \|_{\psi_1} \big), \label{eqn:VWPropA161}\\
\| S_n \|_{\psi_2} &\leq K_2 \big( E|S_n| + (\sum_{i=1}^n \| W_i \|_{\psi_2}^2)^{1/2} \big) \label{eqn:VWPropA162},
\end{align}
where $K_1$ and $K_2$ are constants independent of $n$ and other indices.

2. By Lemma 2.2.2 of VW (2000), for any r.v.'s $W_j$ and $a \geq 1$,
\begin{equation}\label{eqn:VWlemma2.2.2}
\| \max_{1 \leq j \leq m} W_j \|_{\psi_a}  \leq  \tilde{K}_a \max_{1 \leq j \leq m} \|W_j\|_{\psi_a}
(\log(m+1))^{1/a}
\end{equation}
for some constant $\tilde{K}_a$ depending on $a$ only.

3. For any r.v.'s $W_i$ and $a \geq 1$, (see page 105, Q.8 of VW (2000))
\begin{equation}\label{eqn:expectedmaxbound}
E(\max_{1 \leq i \leq m}|W_i|) \leq (\log(m+1))^{1/a} \max_{1 \leq i \leq m} \| W_i \|_{\psi_a}.
\end{equation}

4. For any r.v. $W$ and $a \geq 1$,
\begin{equation}\label{eqn:Orlicznormforsquares}
\| W^2 \|_{\psi_a} = \| W \|_{\psi_{2a}}^2.
\end{equation}

Since the $(y_{jr}\epsilon_{jt})_j$'s are i.i.d. with mean zero (variance bounded by $\sigma_{yM}^2 \sigma_{\epsilon M}^2$ by condition (E)), by (\ref{eqn:VWPropA162}),
\begin{align}
\max_j \| z_{n,j} \|_{\psi_2} &\leq \max_j K_2((Ez_{n,j}^2)^{1/2} + n^{-1/2}(n \|a^2(n)\|_{\psi_2}^2)^{1/2}) \notag \\
&\leq  \max_j K_2(\sigma_{yM} \sigma_{\epsilon M} + O(a^2(n))) = O(a^2(n)). \label{eqn:zbound}
\end{align}
Then using (\ref{eqn:zbound}) and (\ref{eqn:expectedmaxbound}),
\begin{align*}
E(\max_j |z_{n,j}|) &\leq (\log(p_n k_n + 1))^{1/2} \max_j \| z_{n,j} \|_{\psi_2}\\
&=O((\log p_n)^{1/2} a^2(n)),
\end{align*}
which is the inequality (\ref{eqn:expectedzbound}).

\vspace{12pt}
\indent {\bf Step 5.} \; \emph{Proof of (\ref{eqn:expectedetabound}).} \:
By Lemma \ref{lemma:extremeevalues} and condition (B), the eigenvalues $0 < \tau_{t1} \leq \tau_{t2} \leq \cdots \leq \tau_{tk_n} \leq \infty$ of $\C_{t11}$ are uniformly bounded away from $0$ (by $1/\tau$) and $\infty$ (by $\tau$) almost surely when $n \rightarrow \infty$.
Then $\| \C_{t11} \|, \| \C_{t11}^{-1} \| \leq \tau$ almost surely as $n \rightarrow \infty$. Hence for large enough $n$,
$$ \eta_{n,j}^2 = \| \e_k^T \C_{t11}^{-1} \v_{n,t}(1) \|^2 \leq \tau^2 \| \v_{n,t}(1) \|^2, $$
for some $k$ and $t$, where $\e_k$ is the unit vector having the $k$-th position equals to one and zero elsewhere. The vector $\v_{n,t}(1)$ is the truncated version of $\v_t(1)$ containing elements $z_{n,i}$.
Then by (\ref{eqn:Orlicznormforsquares}) and (\ref{eqn:zbound}),
\begin{align}
\max_j \| \eta_{n,j} \|_{\psi_{2}} &= \max_j \| \eta_{n,j}^2 \|_{\psi_1}^{1/2}  \leq \tau \max_{t} \Big\| \| \v_{n,t}(1) \|^2 \Big\|_{\psi_1}^{1/2} \notag \\
&\leq \tau k_n^{1/2} \max_{i=i_1,\cdots,i_{k_n}}\| z_{n,i}^2 \|_{\psi_{1}}^{1/2} = \tau k_n^{1/2} \max_{i=i_1,\cdots,i_{k_n}}\| z_{n,i} \|_{\psi_{2}} \notag \\
&=O( k_n^{1/2} a^2(n)). \label{eqn:etabound}
\end{align}
With this, using (\ref{eqn:expectedmaxbound}), we will arrive at (\ref{eqn:expectedetabound}).

\vspace{12pt}
\indent {\bf Step 6.} \; \emph{Proof of (\ref{eqn:expectedzetabound}).} \:
Since the $ y_{ir}y_{ik}$'s are i.i.d. for each $r$ and $k$  with mean $ \sigma_{rk0} \leq \sigma_{yM}^2$ (variance bounded by $ \sigma_{yM}^4 $ for $r \neq k$), arguments similar to that for (\ref{eqn:zbound}) applies and hence
\begin{equation}\label{eqn:xbound}
\max_{r,k} \| x_{n,rk} \|_{\psi_2} = O(a^2(n)).
\end{equation}
Hence we can use (\ref{eqn:VWlemma2.2.2}), (\ref{eqn:Orlicznormforsquares}), (\ref{eqn:etabound}) and (\ref{eqn:xbound}) to show that
\begin{align}
\max_j \| \zeta_{n,j} \|_{\psi_1} &\leq n^{-1/2} k_n \max_{r,k} \|  \max(x_{n,rk}^2, \eta_{n,k}^2) \|_{\psi_1} \notag \\
&\leq n^{-1/2} k_n \tilde{K}_1 \log 3 \max_{r,k} (\| x_{n,rk} \|_{\psi_{2}}^2, \| \eta_{n,k} \|_{\psi_{2}}^2) \notag \\
&= O( n^{-1/2} k_n^2 a^4(n) ). \label{eqn:zetabound}
\end{align}
With this, using (\ref{eqn:expectedmaxbound}), we will arrive at (\ref{eqn:expectedzetabound}).

\vspace{12pt} {\bf Step 7.} \emph{Proving (\ref{eqn:KKT2}) occurs with probability going to 1 for $k_n=0$.} \; When $k_n = 0$, $\bSigma_0$ is diagonal, and we only need to prove (\ref{eqn:KKT2}) occurs with probability going to 1. Then we need to prove (see Step 3.2 for definition of $x_{kj}$)
$P(\max_{k<j} |x_{kj}| \leq \lambda_n \tilde{w}_{nj}^k / (2n^{1/2})) \rightarrow 1$.

In fact by (\ref{eqn:weightforBZC}), we only need to prove $P(\max_{k<j} |x_{kj}| > \lambda_n n^{1/2} / 2) \rightarrow 0$, which follows from
(\ref{eqn:xbound}) and (\ref{eqn:expectedmaxbound}) and arguments similar to (\ref{eqn:probconv1}) or (\ref{eqn:probconv2}),
\begin{align*}
P(\max_{k<j} |x_{n,kj}| > \lambda_n n^{1/2} / 2) &\leq 2E(\max_{k<j} |x_{n,kj}|) / (\lambda_n n^{1/2}) \\
&= O((\log p_n)^{1/2} a^2(n) / (\lambda_n n^{1/2})) \rightarrow 0,
\end{align*}
by condition (C) and $a(n)$ chosen to go to infinity slow enough. This completes the proof of Theorem \ref{thm:oracleproperty}(i). $\square$

\vspace{12pt}
\noindent {\bf Proof of Theorem \ref{thm:precisionmatrixconsistency}}.
We focus on $\| \hat{\bOmega} - \bOmega_0 \|_{\infty}$ first, which amounts to finding
\begin{equation}\label{eqn:I}
 I = P(\max_{i,j}|\hat{\omega}_{ij} - \omega_{ij0}| > t_n),
\end{equation}
for some $t_n > 0$.

Note that $\omega_{ij} = \sum_{r=1}^{p_n} \sigma_{r0}^{-2} \phi_{r,i} \phi_{r,j}$ with
$ \phi_{i,i} = -1 $ and $ \phi_{i,j} = 0 $ for $ i < j $. We write
$\hat{\omega}_{ij} - \omega_{ij0} = I_1 + \cdots + I_8,$
where ($I_5$ to $I_8$ are omitted since they have orders smaller than
either of $I_1$ to $I_4$ under block sign-consistency)
\begin{align*}
I_1 &= \sum_{k=1}^{p_n} (\hat{\sigma}_k^{-2} - \hat{\sigma}_{k0}^{-2}) \phi_{k,j}^0 \phi_{k,i}^0,
\quad\quad\quad\quad\quad\;\; I_2 =  \sum_{k=1}^{p_n} (\hat{\sigma}_{k0}^{-2} - \sigma_{k0}^{-2}) \phi_{k,j}^0 \phi_{k,i}^0,\\
I_3 &= \sum_{k=1}^{p_n} \sigma_{k0}^{-2} (\hat{\phi}_{k,j} - \phi_{k,j}^0) \phi_{k,i}^0,
\quad\quad\quad\quad\quad\;\; I_4 = \sum_{k=1}^{p_n} \sigma_{k0}^{-2} (\hat{\phi}_{k,i} - \phi_{k,i}^0) \phi_{k,j}^0,%\\
%I_5 &= \sum_{k=1}^{p_n} \sigma_{k0}^{-2} (\hat{\phi}_{k,j} - \phi_{k,j}^0) (\hat{\phi}_{k,i} - \phi_{k,i}^0),
%\quad\quad I_6 = \sum_{k=1}^{p_n} (\hat{\sigma}_k^{-2} - \sigma_{k0}^{-2}) ( \hat{\phi}_{k,i} - \phi_{k,i}^0 ) \phi_{k,j}^0,\\
%I_7 &= \sum_{k=1}^{p_n} (\hat{\sigma}_k^{-2} - \sigma_{k0}^{-2}) ( \hat{\phi}_{k,j} - \phi_{k,j}^0 ) \phi_{k,i}^0,
%\quad\quad I_8 = \sum_{k=1}^{p_n} (\hat{\sigma}_k^{-2} - \sigma_{k0}^{-2}) ( \hat{\phi}_{k,i} - \phi_{k,i}^0 ) ( \hat{\phi}_{k,j} - \phi_{k,j}^0 ),
\end{align*}
and $\hat{\sigma}_{k0}^2 = n^{-1} \sum_{i=1}^n \epsilon_{ik}^2 = n^{-1} \sum_{i=1}^n (y_{ik} - \y_{i[k]}^T \bphi_{k[k]}^0)^2$.
Then, the probability $I$ in (\ref{eqn:I}) can be decomposed as
$$ I \leq \sum_{r=1}^8 a_r P(\max_{i,j}|I_r| > \delta t_n),  $$
where $a_r$ and $\delta$ are absolute constants independent of $n$.

\vspace{12pt} {\bf Step 1.} \emph{Proving the convergence results.} \:
The proof consists of finding the orders of $\max_{i,j}|I_1|$ to $\max_{i,j}|I_4|$. We will show in Step 2 that when $k_n > 0$,
\begin{equation}\label{eqn:maxI3order}
\max_{i,j} |I_{n,3}| = O_P( \{ (k_n+1)^{3} \log p_n / n \}^{1/2}),
\end{equation}
which has the highest order among the four.
When $k_n=0$, $P(I_3 = 0) \rightarrow 1$ by block sign-consistency, and $\max_{i,j}|I_2|$ has order dominating the four. In general, we will show in Step 4 that
\begin{equation}\label{eqn:maxI2order}
\max_{i,j} |I_{n,2}| = O_P( (k_n+1) (\log p_n / n)^{1/2} ).
\end{equation}
Hence
$$ \| \hat{\bOmega} - \bOmega_0 \|_{\infty}^2 =  \max_{i,j} (\hat{\omega}_{ij} - \omega_{ij0})^2 = O_P((k_n+1)^3 \log p_n/n). $$
For $\| \hat{\bOmega} - \bOmega_0 \|$, using the inequality
$\| M \| \leq \max_i \sum_j |m_{ij}|$ for a symmetric matrix M (see e.g. Bickel and Levina (2004)),
we immediately have
$$ \| \hat{\bOmega} - \bOmega_0 \| = O_P((k_n+1)\| \hat{\bOmega} - \bOmega_0 \|_{\infty}), $$
where we used the block sign-consistency and the fact that $\bOmega_0$ has $k_n$ number of non-zero off-diagonals.

\vspace{12pt} {\bf Step 2.} \emph{Proving (\ref{eqn:maxI3order})} \; By the symmetry of $I_3$ and $I_4$, we only need to consider $\max_{i,j}|I_3|$.

\indent \emph{Step 2.1 \: Defining $I_{n,3}$. } \; By block sign-consistency of $\boldsymbol{\phihat}_n$, $\boldsymbol{\hat{\ell}}_1 \cdots \boldsymbol{\hat{\ell}}_{k_n}$ are non-zero with probability going to 1 and (\ref{eqn:KKT1}) is valid for $j=1,\cdots,k_n$. Then we can rewrite (\ref{eqn:KKT1}) into
\begin{equation}\label{eqn:KKT1rewritten}
\C_{t11}(\boldsymbol{\phihat}_{t[t]}(1) - \bphi_{t[t]}^0(1)) = n^{-1/2} \v_t(1)
- \lambda_n \W_{nt} \s_t - \C_{t12} \boldsymbol{\phihat}_{t[t]}(2),
\end{equation}
for $t = 2,\cdots,p_n$. Block sign-consistency implies $\boldsymbol{\phihat}_{t[t]}(2) = 0$ with probability going to 1. Also by (\ref{eqn:weightforBZC}), $\W_{nt} = \0$ with probability going to 1. Hence
\begin{equation*}
\boldsymbol{\phihat}_{t[t]}(1) - \bphi_{t[t]}^0(1) = n^{-1/2} \C_{t11}^{-1} \v_t(1) + o_P(1),
\end{equation*}
where almost sure invertibility of $\C_{t11}$ follows from Lemma \ref{lemma:extremeevalues} and condition (B) as $n \rightarrow \infty$. This implies that, for $j=1,\cdots,k_n$ (note $I_3 = I_4 \equiv 0$ when $k_n=0$) and $t=2,\cdots,p_n$,
\begin{equation}\label{eqn:phihatequiv}
\hat{\phi}_{t,t-j} - \phi_{t,t-j}^0 = n^{-1/2}\eta_k + o_P(1),
\end{equation}
for some $k$, where $\boldeta$ is defined in Step 3.1 in the previous proof. Then we can write $I_3$ as
$$ I_3 = n^{-1/2} \sum_{k=1}^{p_n} \sigma_{k0}^{-2} \eta_{i_k} \phi_{k,i}^0 + o_P(1),  $$
for some intergers $i_1,\cdots, i_{p_n}$. Note that $I_3$ has at most $(k_n+1)$ terms in the above summation. We define
\begin{equation}\label{eqn:In_3}
I_{n,3} = n^{-1/2} \sum_{k=1}^{p_n} \sigma_{k0}^{-2} \eta_{n, i_k} \phi_{k,i}^0,
\end{equation}
where $\eta_{n, i_k}$ is defined in Step 3.1 of the previous proof.

\indent \emph{Step 2.2 \: Finding the order of $\max_{i,j}|I_{3}|$.} \; Under conditions (A) and (E),
$\sigma_{k0}^{-2} \phi_{k,i}^0$ is bounded above uniformly for all $i$ and $k$. Then using (\ref{eqn:etabound}) and (\ref{eqn:expectedmaxbound}),
\begin{align*}
P(\max_{i,j}|I_{n,3}| > \delta t_n) &\leq E(\max_{i,j}|I_{n,3}|)/(\delta t_n)\\
&\leq n^{-1/2} (\log p_n)^{1/2} (k_n+1) \max_{i,j,k} \{ \sigma_{k0}^{-2} \phi_{k,i}^0 \| \eta_{n,i_k} \|_{\psi_2} \} / (\delta t_n)\\
&=O( \{ (k_n+1)^{3} (\log p_n) \}^{1/2} a^2(n) / (n^{1/2} t_n)).
\end{align*}
This shows that $\max_{i,j} |I_{n,3}| = O_P( \{ (k_n+1)^{3} \log p_n / n \}^{1/2})$, which is also the order of $\max_{i,j}|I_3|$, since $\max_{i,j}|I_{n,3} - I_3| \rightarrow 0$ almost surely, and $a(n)$ goes to infinity at arbitrary speed.

\vspace{12pt} {\bf Step 3.} \emph{Showing $I_1 = o_P(I_2)$.} \;
By block sign-consistency, $\boldsymbol{\phihat}_{k[k]}(2) = \0$ with probability going to 1 for $k=2,\cdots,p_n$. Hence
\begin{align*}
\hat{\sigma}_k^2 &= n^{-1} \sum_{i=1}^n (y_{ik} - \y_{i[k]}^T \boldsymbol{\phihat}_{k[k]}(1))^2 + o_P(1) \\
&= \hat{\sigma}_{k0}^2 - 2n^{-1/2} \v_k(1)^T \hat{\u}_{k[k]}(1) + \hat{\u}_{k[k]}(1)^T \C_{k11} \hat{\u}_{k[k]}(1) + o_P(1),
\end{align*}
where $\hat{\u}_{k[k]}(1) = \boldsymbol{\phihat}_{k[k]}(1) - \bphi_{k[k]}^0(1)$. This implies that
\begin{align*}
| \hat{\sigma}_k^2 - \hat{\sigma}_{k0}^2 | &\leq 2n^{-1/2} \| \v_k(1) \| \cdot \| \u_{k[k]}(1) \|
+ \lambda_{\max}(\C_{k11}) \cdot \| \u_{k[k]}(1) \|^2\\
&\leq 2n^{-1/2}O_P(k_n^{1/2}) \cdot O_P(k_n^{1/2} n^{-1/2}) + \tau O_P(k_n/n) = O_P(k_n/n),
\end{align*}
where $\tau$
is an almost sure upper bound for the eigenvalues of $\C_{k11}$ by Lemma \ref{lemma:extremeevalues} and condition (B).
The order for $\| \v_k(1) \|$ can be obtained using ordinary CLT. The order for $\| \hat{\u}_{k[k]}(1) \|$  can be obtained by observing
$ \hat{\phi}_{t,j} - \phi_{t,j}^0 = n^{-1/2} \e_j^T \C_{t11}^{-1} \v_t(1) + o_P(1) $, and by conditioning on $\y_i[t]$ for all $i=1,\cdots,n$,
\begin{align*}
\var(n^{-1/2} \e_j^T \C_{t11}^{-1} \v_t(1)) &= n^{-1}E(\e_j^T \C_{t11}^{-1} \v_t(1) \v_t(1)^T \C_{t11}^{-1} \e_j) \\
&=n^{-1} \sigma_{t0}^2 E(\e_j^T \C_{t11}^{-1} \e_j ) \leq n^{-1} \sigma_{\epsilon M}^2 \tau = O(n^{-1}).
\end{align*}
Hence the delta method shows that
$ \hat{\sigma}_{k}^{-2} - \hat{\sigma}_{k0}^{-2} = O_P(k_n/n) $.

On the other hand,
by the ordinary CLT, we can easily see that $ \hat{\sigma}_{k0}^2 - \sigma_{k0}^2 = O_P(n^{-1/2})$. Thus $I_2$ has a larger order than $I_1$ since $(k_n/n)/n^{-1/2} = k_n n^{-1/2} = o(1)$.
Hence we only need to consider $P(|I_2| > \delta t_n)$ and ignore $P(|I_1| > \delta t_n)$.

\vspace{12pt} {\bf Step 4.} \emph{Proving (\ref{eqn:maxI2order}).} \:
Delta method implies $ \hat{\sigma}_{k0}^{-2} - \sigma_{k0}^{-2} = -\sigma_{k0}^{-4} ( \hat{\sigma}_{k0}^2 - \sigma_{k0}^2 )( 1 + o_P(1) )$.
We then have
\begin{align*}
I_2 &= \sum_{k=1}^{p_n} \bigg\{ -n^{-1} \sum_{r=1}^n
(\epsilon_{rk}^2 - \sigma_{k0}^2) \bigg\} \sigma_{k0}^{-4} \phi_{k,i}^0 \phi_{k,j}^0 (1 + o_P(1)),
\end{align*}
which is a sum of at most $k_n + 1$ terms (corr. $i=j$) of i.i.d. zero mean r.v.'s having uniformly bounded
variance (fourth-moment of $\epsilon_{rk}$) by condition (A). Now define
$$ I_{n,2} = \sum_{k=1}^{p_n} \bigg\{ -n^{-1} \sum_{r=1}^n
(\epsilon_{rk}^2 - \sigma_{k0}^2)\1_{\{ |\epsilon_{rk}^2 - \sigma_{k0}^2| \leq a(n) \}} \bigg\} \sigma_{k0}^{-4} \phi_{k,i}^0 \phi_{k,j}^0, $$
and using (\ref{eqn:expectedmaxbound}) and arguments similar to proving (\ref{eqn:zbound}),
\begin{align*}
P(\max_{i,j} |I_{n,2}| > \delta t_n) &\leq E(\max_{i,j} |I_{n,2}|) / (\delta t_n) \\
&=O( (k_n+1) (\log p_n / n)^{1/2} a(n) / t_n).
\end{align*}
Hence this shows that, by $ \max_{i,j}| I_{n,2} - I_2 | \rightarrow 0 $ almost surely,
$$ \max_{i,j} |I_2| = O_P( (k_n+1) (\log p_n / n)^{1/2}). $$
This completes the proof of the theorem. $\square$

\newpage

\vspace{12pt}
\centerline{\large \bf References}
\begin{description}

\item Antoniadis, A. and Fan, J. (2001). Regularization of wavelets approximations (Disc: p956-967).
{\em J. Amer. Statist. Assoc.}, {\bf 96}, 939-956.

\item Bai, Z. and Silverstein, J.W. (2006), {\em Spectral Analysis
of Large Dimensional Random Matrices,} Science Press, Beijing.

\item Banerjee, O., d'Aspremont, A., and El Ghaoui, L. (2006). Sparse covariance selection via robust
maximum likelihood estimation. In Proceedings of ICML.

\item Bickel, P. J. and Levina, E. (2004). Some theory for Fisher's linear discriminant function, ``naive
Bayes", and some alternatives when there are many more variables than observations. {\em Bernoulli},
{\bf 10(6)}, 989-–1010.

\item Bickel, P. J. and Levina, E. (2007). Covariance Regularization by Thresholding. {\em Ann. Statist.}, to appear.

\item Bickel, P. J. and Levina, E. (2008). Regularized estimation of large covariance matrices. {\em Ann. Statist.}, {\bf 36(1)}, 199--227.

\item Cai, T.T. (1999). Adaptive wavelet estimation: a block thresholding and oracle inequality approach.
{\em Ann. Statist.}, {\bf 27(3)}, 898--924.

\item Cand\`{e}s, E. and Tao, T. (2007). The Dantzig selector: Statistical estimation when $p$ is much larger than $n$.
{\em Ann. Statist.}, {\bf 35(6)}, 2313--2351.

\item Diggle, P. and Verbyla, A. (1998). Nonparametric estimation of covariance structure in longitudinal
data. {\em Biometrics}, {\bf 54(2)}, 401--415.

\item El Karoui, N. (2007). Operator norm consistent estimation of large dimensional
sparse covariance matrices. Technical Report 734, UC Berkeley, Department of Statistics.

\item Fan, J., Fan, Y. and Lv, J. (2007). High dimensional covariance matrix estimation using a factor model. {\em Journal of Econometrics}, to appear.

\item Fan, J., Feng, Y. and Wu, Y. (2007).  Network Exploration
via the Adaptive LASSO and SCAD Penalties.  {\em Manuscript.}

\item Fan, J. and Li, R. (2001). Variable selection via nonconcave penalized
likelihood and its oracle properties. {\em J. Amer. Statist. Assoc.}, {\bf 96}, 1348-–1360.

\item Fan, J. and Zhang, W. (2000). Statistical estimation in varying coefficient models. {\em Ann. Statist.}, {\bf 27}, 1491--1518.

\item Friedman, J., Hastie, T., and Tibshirani, R. (2007). Pathwise coordinate optimization. Technical
report, Stanford University, Department of Statistics.

\item Furrer, R. and Bengtsson, T. (2007). Estimation of high-dimensional prior and posterior covariance
matrices in Kalman filter variants. {\em Journal of Multivariate Analysis}, {\bf 98(2)}, 227-–255.

\item Graybill, F.A. (2001), {\em Matrices with Applications in
Statistics (2nd ed.),} Belmont, CA: Duxbury Press.

\item Huang, J., Liu, N., Pourahmadi, M., and Liu, L. (2006). Covariance matrix selection and estimation
via penalised normal likelihood. {\em Biometrika}, {\bf 93(1)}, 85-–98.

\item Huang, J., Ma, S. and Zhang, C.H. (2006). Adaptive LASSO for sparse high-dimensional regression
models. Technical Report {\bf 374}, Dept. of Stat. and Actuarial Sci., Univ. of Iowa.

\item Kim, Y., Kim, J. and Kim, Y. (2006). Blockwise sparse regression. {\em Statist. Sinica}, {\bf 16}, 375--390.

\item Lam, C. and Fan, J. (2007). Sparsistency and rates of convergence in large covariance matrices estimation. {\em Manuscript}.

\item Levina, E., Rothman, A.J. and Zhu, J. (2007). Sparse Estimation of Large Covariance
Matrices via a Nested Lasso Penalty, {\em Ann. Applied Statist.}, to appear.

\item Li, H. and Gui, J. (2006). Gradient directed regularization for sparse Gaussian concentration graphs, with
applications to inference of genetic networks. {\em Biostatistics} {\bf 7(2)}, 302--317.

\item Mar\^{c}enko, V.A. and Pastur, L.A. (1967).  Distributions of eigenvalues of some sets of random matrices.  {\em Math. USSR-Sb},
{\bf 1}, 507--536.

\item Meinshausen, N. and Buhlmann, P. (2006). High dimensional graphs and variable selection with the Lasso.
{\em Ann. Statist.}, {\bf 34}, 1436-–1462.

\item Pourahmadi, M. (1999). Joint mean-covariance models with applications to longitudinal data:
unconstrained parameterisation. {\em Biometrika}, {\bf 86}, 677-–690.

\item Rothman, A.J., Bickel, P.J., Levina, E., and Zhu, J. (2007).  Sparse Permutation Invariant Covariance Estimation. Technical report  {\bf 467},  Dept. of Statistics, Univ. of Michigan.

\item Rothman, A.J., Levina, E. and Zhu, J. (2008). Generalized Thresholding of Large Covariance Matrices. Technical report, Dept. of Statistics, Univ. of Michigan.

\item Shen, H. and Huang, J. Z. (2005). Analysis of call center data using singular value decomposition.
{\em App. Stochastic Models in Busin. and Industry}, {\bf 21}, 251--263.

\item Smith, M. and Kohn, R. (2002). Parsimonious covariance matrix estimation for longitudinal data.
{\em J. Amer. Statist. Assoc.}, {\bf 97(460)}, 1141-–1153.

\item van der Vaart, A.W. and Wellner, J.A. (2000). {\em Weak Convergence and Empirical Processes: With Applications to Statistics}. Springer, New York.

\item Wagaman, A.S. and  Levina, E. (2007). Discovering Sparse Covariance Structures with the Isomap. Technical report  {\bf 472},  Dept. of Statistics, Univ. of Michigan.

\item Wu, W. B. and Pourahmadi, M. (2003). Nonparametric estimation of large covariance matrices of
longitudinal data. {\em Biometrika}, {\bf 90}, 831-–844.

\item Yuan, M. and Lin, Y. (2006). Model selection and estimation in regression with
grouped variables. {\em J. R. Stat. Soc. B}, {\bf 68}, 49-–67.

\item Yuan, M. and Lin, Y. (2007). Model selection and estimation in the Gaussian graphical model.
{\em Biometrika}, {\bf 94(1)}, 19–-35.

\item Zhao, P., Rocha, G. and Yu, B. (2006). Grouped and hierarchical
model selection through composite absolute penalties. {\em Ann. Statist.}, to appear.

\item Zhao, P. and Yu, B. (2006). On model selection consistency of lasso. Technical Report, Statistics Department, UC Berkeley.

\item Zou, H. (2006). The Adaptive Lasso and its Oracle Properties. {\em J. Amer. Statist. Assoc.}, {\bf 101(476)}, 1418-–1429.

\item Zou, H. and Li, R. (2008). One-step sparse estimates in nonconcave penalized likelihood models. {\em Ann. Statist.}, to appear.

\end{description}

\newpage
\centerline{\large \bf Supplement: Proof of Theorems \ref{thm:oracleproperty}(ii) and \ref{thm:initialestimatorBZC}}
\setcounter{equation}{0}
\renewcommand \theequation{S.\arabic{equation}}

\noindent {\bf Proof of Theorem \ref{thm:oracleproperty}(ii)}.
To prove asymptotic normality for $\boldsymbol{\phihat}_{n1}$, note that by (\ref{eqn:KKT1rewritten}), for $\balpha_n$ with $\|
\balpha_n \| = 1$ and $\nu_n = \balpha_n \H_n \balpha_n$,
\begin{equation}\label{eqn:unhat}
n^{1/2} \nu_n^{-1/2} \balpha_n^T (\boldsymbol{\phihat}_{n1} - \bphi_{n1}^0) = I_1 + I_2 +
I_3,
\end{equation}
where  $ I_2 = \lambda_n
(n\nu_n)^{-1/2} \balpha_n^T \G_{11}^{-1} \W_n \s /2 $\; , \;$ I_3 =
(n/\nu_n)^{1/2} \balpha_n^T \G_{11}^{-1} \G_{12} \boldsymbol{\phihat}_{n2}$ and $I_1 = \nu_n^{-1/2} \balpha_n^T \G_{11}^{-1} \z$, with
$\bphi_{n2}$ the vector of elements of $\bphi_n$ corresponding to
its zero off-diagonals.

\vspace{12pt} {\bf Step 1.} \emph{Showing $I_2, I_3 = o_P(1)$. } \:
Since $P(\boldsymbol{\phihat}_{n2} = \0) \rightarrow 1$, we have $P(I_3 =
0) \rightarrow 1$, thus $I_3 = o_P(1)$. Also, we can easily show that
$$ |I_2| \leq C \tau_1^{-1} a_n (nl_n)^{1/2} \nu_n^{-1/2} k_n /2,  $$
where $a_n = \max \{ p_{\lambda_{nj}}^\prime (\| \bl_j^{(k)} \|) :
j \in J_{n1} \}$. Hence if $a_n = o(\nu_n^{1/2} (nl_n)^{-1/2} k_n^{-1})$,
we have $|I_2| = o_P(1)$. The SCAD penalty ensures that $a_n = 0$ for sufficiently large $n$
if the initial estimator $\bphi_n^{(k)}$ is good enough, which is measured by its block zero-consistency.

\vspace{12pt} {\bf Step 2.}
We write $\balpha_n = (\balpha_{n2}^T, \cdots, \balpha_{np_n})^T$, so that
$I_1 = \nu_n^{-1/2} \sum_{j=2}^{p_n} \balpha_{nj}^T \C_{j11}^{-1} \v_j(1)$. Define
$$ \tilde{I}_1 = \nu_n^{-1/2} \sum_{j=2}^{p_n} \balpha_{nj}^T \Sigma_{j11}^{-1} \v_j(1), $$
where $\Sigma_{j11} = E(\C_{j11})$. We can rewrite $\tilde{I}_1 = \sum_{i=1}^n w_{n,i} $, where
$$ w_{n,i} = (n\nu_n)^{-1/2} \sum_{j=2}^{p_n} \balpha_{nj}^T \Sigma_{j11}^{-1} \epsilon_{ij} \y_{i[j]}(1)  $$
are independent and identically distributed with mean zero for all $i$. Our aim is to utilize the Lindeberg-Feller CLT
to prove asymptotic normality of $\tilde{I}_1$, then argue that $I_1$ itself is distributed like $\tilde{I}_1$, thus
finishing the proof.

\vspace{12pt} {\bf Step 3.} \emph{Showing asymptotic normality for $\tilde{I}_1$.} \:
First, by suitably conditioning on the filtration $\mathcal{F}_{t} = \sigma\{\bepsilon_1, \cdots, \bepsilon_t\}$ generated by
the $\bepsilon_j = (\epsilon_{1j},\cdots,\epsilon_{nj})^T$ for $j=1,\cdots,t$, we can show that (proof omitted) $\var (\tilde{I}_1) = 1$.

\indent \emph{Step 3.1 \: Checking the Lindeberg's condition.} \;
Next, by the Cauchy-Schwarz inequality, for a fixed $\epsilon > 0$,
\begin{align*}
\sum_{i=1}^n Ew_{n,i}^2 \1_{\{ |w_{n,i}|>\epsilon \}} &= nE(w_{n,1}^2 \1_{\{ |w_{n,1}|>\epsilon \}}) \\
&\leq \nu_n^{-1} \Bigg\{ E\Big(\sum_{j=2}^{p_n} \balpha_{nj}^T \Sigma_{j11}^{-1} \epsilon_{1j} \y_{1[j]}(1) \Big)^4 \Bigg\}^{1/2} \cdot
\{ P(w_{n,1}^2 > \epsilon^2) \}^{1/2}.
\end{align*}
\indent \emph{Step 3.1.1} \; The Markov inequality implies that
$$ P(w_{n,1}^2 > \epsilon^2) <  \epsilon^{-2}E(w_{n,1}^2) = \epsilon^{-2}n^{-1},$$ thus $\{ P(w_{n,1}^2 > \epsilon^2) \}^{1/2} = O(n^{-1/2})$.

\indent \emph{Step 3.1.2} \; For the former expectation, note that condition (B) implies that the eigenvalues of $\Sigma_{j11}$ are uniformly bounded away from zero and
infinity as well, say by $c^{-1}$ and $c$ respectively, so that $\| \Sigma_{j11}^{-1} \| \leq c$ for all $j$. Hence
\begin{align*}
E\Big( \sum_{j=2}^{p_n} \balpha_{nj} \Sigma_{j11}^{-1} \epsilon_{1j} \y_{1[j]}(1) \Big)^4
&\leq c^4E(\max_{j} |\epsilon_{1j}| \|\y_{1[j]}(1)\|)^4 \cdot \big( \sum_{j=2}^{p_n} \| \balpha_{nj} \| \big)^4 \\
&\leq c^4 k_n^2 E(\max_{j:\balpha_{nj} \neq 0} |\epsilon_{1j}| \|\y_{1[j]}(1)\|)^4\\
&\leq c^4 k_n^2 E(\max_{j:\balpha_{nj} \neq 0} \epsilon_{1j}^4) \cdot E(\| \y_{1[p_n]}(1) \|^4),
\end{align*}
where the second line used the fact that there are at most $k_n$ of the $\balpha_{nj}$ that are non-zero and that
$\sum_{j=2}^{p_n} \| \balpha_{nj} \|^2 = 1$ implies $\big( \sum_{j=2}^{p_n} \| \balpha_{nj} \| \big)^4 \leq k_n^2$. The third line used
conditioning arguments and the fact that $\y_{1[p_n]}(1)$ has the largest magnitude among the $\y_{1[j]}(1)$'s.  With the tail assumptions for
the $\epsilon_{ij}$'s and the $y_{ij}$'s in condition (A), the fourth moments for $\max_{j:\balpha_{nj} \neq 0} \epsilon_{1j}$ and
$\| \y_{1[p_n]}(1) \|$ exist. Using (\ref{eqn:VWlemma2.2.2}) and (\ref{eqn:expectedmaxbound}), can show
$$E(\max_{j:\balpha_{nj} \neq 0} \epsilon_{1j}^4) = O(\{ \log(k_n+1) \}^{4/d}), \;\; E(\| \y_{1[p_n]}(1) \|^4) = O(k_n^2(\log(k_n+1))^{4/d}).$$
Hence $ E\Big(\sum_{j=2}^{p_n} \balpha_{nj}^T \Sigma_{j11}^{-1} \epsilon_{1j} \y_{1[j]}(1) \Big)^4 = O(k_n^4(\log^2 (k_n+1))^{4/d})$, and
combining previous results we have
$$ \sum_{i=1}^n Ew_{n,i}^2 \1_{\{ |w_{n,i}|>\epsilon \}} = O(k_n^2 (\log(k_n+1))^{4/d} n^{-1/2}\nu_n^{-1})=o(1)  $$
by our assumption stated in the theorem. Hence Lindeberg-Feller CLT implies that $\tilde{I}_1 \dcv N(0,1)$.

\vspace{12pt} {\bf Step 4.} \emph{Showing $I_1$ is distributed similar to $\tilde{I}_1$.} \:
Finally, note that $E(I_1 - \tilde{I}_1) = 0$ and using conditioning arguments as before, we have
\begin{align*}
\var(I_1 - \tilde{I}_1) &= \sum_{j=2}^{p_n}
\sigma_{j0}^2 E(\balpha_{nj}^T(\C_{j11}^{-1} - \Sigma_{j11}^{-1}) \C_{j11} (\C_{j11}^{-1} - \Sigma_{j11}^{-1}) \balpha_{nj})\\
&\leq \max_{1 \leq j \leq p_n} \sigma_{j0}^2 E(\| \C_{j11}^{-1} - \Sigma_{j11}^{-1} \|^2 \cdot \| \C_{j11} \|)\\
&\leq \max_{1 \leq j \leq p_n} \sigma_{j0}^2 E(\| \Sigma_{j11}^{-1} \|^2 \cdot \| \Sigma_{j11} \|^2 \cdot
\| \Sigma_{j11}^{-1/2} \C_{j11} \Sigma_{j11}^{-1/2} - I \|^2 \cdot \| \C_{j11}^{-1} \|^2 \cdot \| \C_{j11} \|).
\end{align*}
As discussed before, we have $\| \Sigma_{j11} \| \leq c$ and $\| \Sigma_{j11}^{-1} \| \leq c$. Also, the semicircular law
implies that $\| \Sigma_{j11}^{-1/2} \C_{j11} \Sigma_{j11}^{-1/2} - I \|^2 = O_P(k_n/n)$. We also have, almost surely, $\|\C_{j11}\|$, $\|\C_{j11}\| \leq \tau$ for each $j=2,\cdots,p_n$ as $n \rightarrow \infty$. Hence for large enough $n$, by condition (E),
$$ \var(I_1 - \tilde{I}_1) \leq c^4\tau^2 \max_{1 \leq j \leq p_n} \sigma_{j0}^2 \cdot O(k_n/n) = o(1),$$
so that $I_1 = \tilde{I}_1 + o_P(1)$, and this completes the proof. $\square$

\vspace{12pt} \noindent {\bf Proof of Theorem
\ref{thm:initialestimatorBZC}}.
The true model for $\y_i = (y_{1i}, \cdots, y_{ni})^T$
(refer to (\ref{eqn:initialestimator})) is
\begin{equation}\label{eqn:truemodel}
\y_i = \tilde{\X}_{i1} \bphi_{i[i]1}^0 + \bepsilon_i,
\end{equation}
for $i=2,\cdots,p_n$, where (recall that $c_{ni} = \max(\lfloor i - \gamma n \rfloor, 1)$
$$ \tilde{\X}_i = (\y_{c_{ni}}, \cdots, \y_{i-1}), \quad \bphi_{i[i]1} = (\phi_{i,c_{ni}},\cdots,\phi_{i,i-1})^T. $$
%Recursively, with $c_{ni}^{(1)} = c_{ni}$, and  $ c_{ni}^{(r)} = \max(\lfloor c_{ni}^{(r-1)} - \gamma n \rfloor, 1) $ for $r \geq 2$, we define
%$$ \tilde{\X}_{ir} = (\y_{c_{ni}^{(r)}},\cdots,\y_{c_{ni}^{(r-1)} - 1}),
%\quad \bphi_{i[i]r} = (\phi_{i,c_{ni}^{(r)}},\cdots,\phi_{i,c_{ni}^{(r-1)}-1})^T. $$

\vspace{12pt} {\bf Step 1.} \emph{To show $P(\max_{j \in J_{n0}} \|\boldsymbol{\tilde{\ell}}_j\|/(p_n-j)^{1/2} \geq \gamma_n) \rightarrow 0$.} \:
We need the following results, the first of which will be proved in Step 3: For each $j \in J_{n0}$ with $1 \leq j \leq \lfloor \gamma n \rfloor$,
\begin{equation}\label{eqn:ellj4thmoment}
E(\| \boldsymbol{\tilde{\ell}}_j \|^4 / (p_n - j)^2) = O(n^{-2}),
\end{equation}
and, for a non-decreasing convex function $\psi$ with $\psi(0) = 0$, a generalization of (\ref{eqn:expectedmaxbound}),
\begin{equation}\label{eqn:generalexpectedmaxbound}
E(\max_{1 \leq i \leq m} |W_i|) \leq \psi^{-1} (m) \max_{1 \leq i \leq m} \| W_i \|_{\psi}.
\end{equation}
Then, with the function $\psi(x) = x^4$ in (\ref{eqn:generalexpectedmaxbound}), using (\ref{eqn:ellj4thmoment}), and $\gamma_n > 0$,
\begin{align*}
P(\max_{j \in J_{n0}} \|\boldsymbol{\tilde{\ell}}_j\|/(p_n-j)^{1/2} \geq \gamma_n)
&\leq E(\max_{j \in J_{n0}} \|\boldsymbol{\tilde{\ell}}_j\|^4 / (p_n-j)^{2}) / \gamma_n^4 \\
&= E(\max_{j \in J_{n0}, 1 \leq j \leq \lfloor \gamma n \rfloor} \|\boldsymbol{\tilde{\ell}}_j\|^4 / (p_n-j)^{2}) / \gamma_n^4\\
&\leq (\lfloor \gamma n \rfloor)^{1/4} \max_{j \in J_{n0}, 1 \leq j \leq \lfloor \gamma n \rfloor} \{ E(\| \boldsymbol{\tilde{\ell}}_j \|^4 / (p_n - j)^2) \}^{1/4}\\
&=O(n^{-1/4}) \rightarrow 0,
\end{align*}
where the second line used the fact that we have set the off-diagonal bands more than $\lfloor \gamma n \rfloor$ bands from the main diagonal to zero.

\vspace{12pt} {\bf Step 2.} \: \emph{To show $P(\min_{j \in J_{n1}} \|\boldsymbol{\tilde{\ell}}_j\|/(p_n-j)^{1/2} \geq \gamma_n) \rightarrow 1$.} \:
We need the following result, which will be proved in Step 4: For $j \in J_{n1}$,
\begin{equation}\label{eqn:ellj2ndmoment}
E(\| \boldsymbol{\tilde{\ell}}_j \|^2 / (p_n - j)) = \| \boldsymbol{{\ell}}_{j0} \|^2 / (p_n - j) + O(n^{-1}).
\end{equation}
Then with $\gamma_n < \min_{j \in J_{n1}} \| \boldsymbol{{\ell}}_{j0} \| / (p_n - j)^{1/2}$,
writing $a_j = (\gamma_n - \| \boldsymbol{{\ell}}_{j0} \| / (p_n - j)^{1/2} )^2$,
\begin{align*}
P(\min_{j \in J_{n1}} \|\boldsymbol{\tilde{\ell}}_j\|/&(p_n-j)^{1/2} \geq \gamma_n)
\geq 1 - \sum_{j \in J_{n1}} P(\|\boldsymbol{\tilde{\ell}}_j\|/(p_n-j)^{1/2} \leq \gamma_n)\\
&\geq 1 - \sum_{j \in J_{n1}} P \Big( (\|\boldsymbol{\tilde{\ell}}_j\| - \|\boldsymbol{{\ell}}_{j0}\|)^2 / (p_n-j) \geq (\gamma_n - \| \boldsymbol{{\ell}}_{j0} \| / (p_n - j)^{1/2} )^2 \Big)\\
%&\geq 1 - \sum_{j \in J_{n1}} a_j^{-1} (p_n - j)^{-1} \{ 2 \| \boldsymbol{\ell}_{j0} \|^2 + O(n^{-1}(p_n - j))
%- 2\| \boldsymbol{\ell}_{j0} \| \cdot  \}\\
&\approx 1 - \sum_{j \in J_{n1}} 2 a_j^{-1} (p_n - j)^{-1} \| \boldsymbol{\ell}_{j0} \|^2 \{ 1 - (1 + O(n^{-1}(p_n-j)))^{1/2} + O(n^{-1}(p_n-j)) \}\\
&= 1 - O(k_n / n) \rightarrow 1,
\end{align*}
where the second last line used the delta method, with (\ref{eqn:ellj4thmoment}) showing the remainder term is going to zero. From Steps 1 and 2, we need to choose $ 0 < \gamma_n < \min_{j \in J_{n1}} \|\boldsymbol{{\ell}}_{j0}\| / (p_n-j)^{1/2}$.

\vspace{12pt} {\bf Step 3.} \: \emph{To prove (\ref{eqn:ellj4thmoment}).} \:
We need the following result, which can be easily generalized from Theorems 10.9.1, 10.9.2 and 10.9.10(1) of Graybill (2001):
Let $\bepsilon = (\epsilon_1, \cdots, \epsilon_m)^T$, where the $\epsilon_i$'s are i.i.d. with mean $0$, and with finite second and fourth moments. Then for symmetric constant matrices $A$ and $B$,
\begin{equation}\label{eqn:matrix4thmomentformula}
E((\bepsilon^T A \bepsilon)(\bepsilon^T B \bepsilon)) = a\tr(A)\tr(B) + b\tr(AB),
\end{equation}
where $a$ and $b$ are constants depending on the second and fourth moments of $\epsilon_i$ only.

The estimator $\tilde{\T}$, obtained from a series of linear regressions introduced in the theorem, has rows such that
by (\ref{eqn:truemodel}),
\begin{align*}
\boldsymbol{\tilde{\phi}}_{i[i]1} &= (\tilde{\X}_{i}^T \tilde{\X}_{i})^{-1} \tilde{\X}_{i}^T \y_i.
\end{align*}
Using (\ref{eqn:truemodel}), for $j \in J_{n0}$ and $1 \leq j \leq \lfloor \gamma n \rfloor$, it is easy to see that
\begin{align*}
\| \boldsymbol{\tilde{\ell}}_{j} \|^2 / (p_n - j) &= (p_n - j)^{-1} \sum_{i=j+1}^{p_n} (\e_{r_{i,j}}^T (\tilde{\X}_i^T \tilde{\X}_i)^{-1} \tilde{\X}_i^T \bepsilon_i)^2 \\
&=(p_n-j)^{-1} \sum_{i=j+1}^{p_n} \bepsilon_i^T A_i \bepsilon_i,
\end{align*}
where $ A_i = \tilde{\X}_i (\tilde{\X}_i^T \tilde{\X}_i)^{-1} \e_{r_{i,j}} \e_{r_{i,j}}^T (\tilde{\X}_i^T \tilde{\X}_i)^{-1} \tilde{\X}_i^T $, and $r_{i,j}$ is some constant depending on $i$ and $j$. With this notation, we have
\begin{align*}
\| \boldsymbol{\tilde{\ell}}_j \|^4 / (p_n-j)^2 = (p_n-j)^{-2} \sum_{r,k=j+1}^{p_n} (\bepsilon_r^T A_r \bepsilon_r)(\bepsilon_k^T A_k \bepsilon_k).
\end{align*}
It is then sufficient to show that $E((\bepsilon_r^T A_r \bepsilon_r)(\bepsilon_k^T A_k \bepsilon_k)) = O(n^{-2})$ for each $r \geq k$.
Let $\mathcal{F}_{i-1} = \sigma\{ \bepsilon_1,\cdots,\bepsilon_{i-1} \}$ be the sigma algebra generated by the $\bepsilon_k$ for $1 \leq k \leq i-1$.
For large enough $n$, we have by Lemma \ref{lemma:extremeevalues} and condition (B), for some constant $B_{\gamma}$ independent of $n$, and for each $i = j+1,\cdots, p_n$,
\begin{equation}\label{eqn:tracebound}
\tr(A_i) = \e_{r_{i,j}}^T (\tilde{\X}_i^T \tilde{\X}_i)^{-1} \e_{r_{i,j}} \leq B_{\gamma}n^{-1}.
\end{equation}
\indent \emph{Step 3.1 \: To show $E((\bepsilon_r^T A_r \bepsilon_r)(\bepsilon_k^T A_k \bepsilon_k)) = O(n^{-2})$ for $r > k$.} \:
Hence for $r > k$ with large enough $n$, using (\ref{eqn:tracebound}),
\begin{align*}
E((\bepsilon_r^T A_r \bepsilon_r)(\bepsilon_k^T A_k \bepsilon_k)) &= E(\bepsilon_k^T A_k \bepsilon_k E_{\mathcal{F}_{r-1}}(\bepsilon_r^T A_r \bepsilon_r)) = E(\bepsilon_k^T A_k \bepsilon_k \sigma_{r0}^2 \tr(A_r)) \\
&\leq B_{\gamma} \sigma_{\epsilon M}^2 n^{-1} E(\bepsilon_k^T A_k \bepsilon_k) = B_{\gamma} \sigma_{\epsilon M}^2 n^{-1} E(\sigma_{k0}^2 \tr(A_k))\\
&\leq B_{\gamma}^2 \sigma_{\epsilon M}^4 n^{-2} = O(n^{-2}).
\end{align*}

\indent \emph{Step 3.2 \: To show $E((\bepsilon_r^T A_r \bepsilon_r)^2) = O(n^{-2})$.} \:
Using (\ref{eqn:matrix4thmomentformula}), with constants $a$ and $b$ uniformly bounded by condition (A) and condition (E), it is sufficient to show that for large enough $n$, $\tr^2(A_r)$ and $\tr(A_r^2)$ are $O(n^{-2})$. By (\ref{eqn:tracebound}) we have $\tr^2(A_r) = O(n^{-2})$. Also,
\begin{align*}
\tr(A_r^2) = (\e_{r_{i,j}}^T (\tilde{\X}_r^T \tilde{\X}_r)^{-1} \e_{r_{i,j}})^2 \leq B_{\gamma}^2 n^{-2},
\end{align*}
for large enough $n$, by (\ref{eqn:tracebound}).

\vspace{12pt} {\bf Step 4.} \: \emph{To prove (\ref{eqn:ellj2ndmoment}).} \:
For $j \in J_{n1}$ and large enough $n$,
\begin{align*}
E(\| \boldsymbol{\tilde{\ell}}_j \|^2 / (p_n - j)) &= \| \boldsymbol{\ell}_{j0} \|^2 / (p_n - j)
+ (p_n-j)^{-1} \sum_{i=j+1}^{p_n} E(\bepsilon_i^T A_i \bepsilon_i)\\
&\leq \| \boldsymbol{\ell}_{j0} \|^2 / (p_n - j) + \sigma_{\epsilon M}^2 \max_i E(\tr(A_i))\\
&\leq \| \boldsymbol{\ell}_{j0} \|^2 / (p_n - j) + O(n^{-1}),
\end{align*}
where the last line used (\ref{eqn:tracebound}). This completes the proof of the theorem. $\square$

\end{document}